%% file: Three_gen_models.tex
\documentclass[11pt]{article}

\usepackage{bbm}
\usepackage{diagrams}
\usepackage{graphicx}
\usepackage{fullpage}
\usepackage{hyperref}
\usepackage{color}

\input{macros.tex}

\def\str{\vrule height16pt width0pt depth10pt}
\newcommand{\Xt}{\widetilde X}
\newcommand{\Vt}{\widetilde V}
\newcommand{\Ft}{
 \cF}
\newcommand{\Gt}{
 \cG}

\numberwithin{equation}{section}

\begin{document}
\setcounter{page}{0}
\pagestyle{empty}

\begin{center}
{\LARGE The MSSM Spectrum from (0,\,2)-Deformations of the\\Heterotic Standard Embedding\\}
\vspace{.7in}
        Volker Braun$^*$\footnote{\it vbraun@stp.dias.ie}, 
        Philip Candelas$^\dagger$\footnote{\it candelas@maths.ox.ac.uk},\\[2ex]
        Rhys Davies$^\dagger$\footnote{\it daviesr@maths.ox.ac.uk} and
        Ron Donagi$^\ddag$\footnote{\it donagi@math.upenn.edu}\\
\vspace{.4in}
{\it
${}^*$Dublin Institute for Advanced Studies\\
10 Burlington Road\\
Dublin 4, Ireland}\\
\vspace{.2in}

{\it
\begin{tabular}{cc}
${}^\dagger$Mathematical Institute, & ${}^\ddag$Department of Mathematics,\\
University of Oxford, & University of Pennsylvania,\\
24-29 St Giles, Oxford & Philadelphia \\
OX1 3LB, UK &  PA 19104-6395, USA
\end{tabular}
}
\end{center}
\vspace{.2in}

\begin{abstract}
We construct supersymmetric compactifications of $E_8{\times}E_8$
heterotic string theory which realise exactly the massless spectrum of the
Minimal Supersymmetric Standard Model (MSSM) at low energies.  The
starting point is the standard embedding on a Calabi-Yau threefold which
has Hodge numbers $\hodgenos = (1,4)$ and fundamental group $\IZ_{12}$,
which gives an $E_6$ grand unified theory with three net chiral generations.
The gauge symmetry is then broken to that of the standard model by a
combination of discrete Wilson lines and continuous deformation of the gauge
bundle.  On eight distinct branches of the moduli space, we find stable bundles
with appropriate cohomology groups to give exactly the massless spectrum of
the MSSM.
\end{abstract}

\setcounter{footnote}{0}
\newpage
\pagestyle{plain}

\tableofcontents

\setlength{\parskip}{.5ex plus .2ex minus .2ex}
\addtolength{\skip\footins}{2ex}

\vspace{-2ex}
\section{Introduction}\label{sec:intro}

The first attempts at finding realistic particle physics in string theory were
based on `standard embedding' compactifications of the
$E_8{\times}E_8$ heterotic string \cite{Candelas:1985en}.  The standard
embedding is a general solution, in which spacetime is taken to be a direct
product of Minkowski space $\cM_4$ and a Calabi-Yau threefold $X$, and
the gauge fields corresponding to one $E_8$ are set `equal' to the
Levi-Civita connection on $X$ (which takes values in $\mathfrak{su}(3)$)
via the embedding
$\mathfrak{su}(3)\subset \mathfrak{su}(3){\times}\mathfrak{e}_6 \subset
\mathfrak{e}_8$.  The gauge fields of the other $E_8$ are set to zero.
This leads to an $\cN = 1$ supersymmetric $E_6$ GUT, with $h^{2,1}(X)$
and $h^{1,1}(X)$ chiral multiplets in the $\rep{27}$ and $\conjrep{27}$
representations respectively.  In order to obtain three net generations of
chiral fermions, then, one must find a Calabi-Yau threefold with Euler
number satisfying $\frac 12 \chi(X) = h^{1,1} - h^{2,1} = \pm 3$.

Breaking of $E_6$ to the standard model gauge group
$\GSM = SU(3){\times}SU(2){\times}U(1)$ can then be achieved by a
combination of discrete Wilson lines and a four-dimensional
supersymmetric Higgs mechanism, which also gives mass to some of
the extraneous matter \cite{Greene:1986jb}.  The vacuum expectation
values (VEVs) are required to be very large, and this makes it hard to
keep control of the theory.  For example, the light spectrum after
Higgsing depends heavily on not only the Yukawa couplings of the
original theory, but also non-renormalisable terms, and these are unlikely
to be readily calculable.

For these reasons, among others, heterotic model building has moved
away from the standard embedding to more general backgrounds, in
which the geometry is still Calabi-Yau, but the gauge fields are given by
other solutions of the Hermitian-Yang-Mills equations.  Such solutions
correspond to slope-stable holomorphic vector bundles on $X$
\cite{Donaldson:1985zz,UhlenbeckYau}.  Although establishing the
stability of bundles is a difficult problem (see e.g.
\cite{Braun:2006ae,Anderson:2008uw,Anderson:2009sw,Anderson:2009nt}),
this approach has led to the discovery of models with the spectrum of the
MSSM \cite{Bouchard:2005ag,Anderson:2011ns}, as well as extensions
thereof which include a massless gauge boson coupling to baryon
number minus lepton number\footnote{Such models were also pursued
in \cite{Braun:2005nv}, but in that example, anomaly cancellation requires
the introduction of anti-branes, and the compactification therefore breaks
supersymmetry (additional examples like this appear in
\cite{Bouchard:2008bg}).} \cite{Anderson:2009mh}.

In this paper we construct the first models which realise exactly the massless
spectrum of the MSSM via deformation of the standard embedding---such models
correspond to (0,\,2) rank-changing deformations of the worldsheet theory
\cite{McOrist:2011bn}.  The compactification manifold is the recently-discovered
Calabi-Yau threefold with Hodge numbers $\hodgenos = (1,4)$ and
fundamental group $\IZ_{12}$ \cite{Braun:2009qy,Braun:2010vc}.  The gauge bundles are irreducible
$SU(5)$ bundles obtained as deformations of the standard embedding
solution, thus preserving the characteristic classes and ensuring that we
retain three chiral generations.  The technique used to construct the
deformed bundles is the one introduced in \cite{Li:2004hx}, and also applied
in \cite{Donagi:2006yf}, with the added complication of retaining invariance
under the quotient group $\IZ_{12}$. Continuous deformation of the gauge
bundle corresponds to the supersymmetric Higgs mechanism in the low
energy theory i.e. giving VEVs to charged flat directions, but constructing
the bundle directly ensures that we do in fact have a consistent
compactification, and also allows a direct and reliable calculation of the
spectrum.

Compactifying on one of these stable rank-five bundles leaves
$SU(5)_\GUT$ as the unbroken gauge group, and this can then be broken
to $\GSM$ by discrete Wilson lines.  This two step approach is necessary,
as the additional adjoint fields required to Higgs $SU(5)_\GUT$ to $\GSM$
are not present in heterotic compactifications, and $E_6$ cannot be broken
directly to the $\GSM$ by any choice of discrete Wilson lines
\cite{McInnes:1989rg}.

Before embarking on our long journey through the technical details, we will
summarise the main ideas and results here.  We start with a Calabi-Yau
manifold $\Xt$ which has Hodge numbers $\hodgenos = (8,44)$ and can be
realised as a hypersurface inside a toric fourfold.  For special choices of its
complex structure, $\Xt$ admits a free action by the group $\IZ_{12}$, such
that the quotient manifold $X = \Xt/\IZ_{12}$
has Hodge numbers $\hodgenos = (1,4)$.  The standard embedding on $X$
would therefore yield an $E_6$ GUT with three generations of particles in
the $\rep{27}$ representation, and one extra vector-like generation.
Note that in a previous paper \cite{Braun:2009qy}, attention was focussed on
a quotient of $\Xt$ by the non-Abelian group $\Dic_3 = \IZ_3\rtimes \IZ_4$,
which also yields Hodge numbers $\hodgenos = (1,4)$.  Unfortunately,
using the methods described here, we subsequently discovered that this
manifold does not admit models with the MSSM spectrum.  This result is
explained briefly in \aref{app:Dic3_models}.

Our approach will be to treat the standard embedding as corresponding to
the degenerate rank-five bundle $T\Xt{\oplus}\cO_{\Xt}{\oplus}\cO_{\Xt}$,
where $T\Xt$ is the tangent bundle, and $\cO_{\Xt}$ is the trivial line
bundle, and then deform this to an irreducible rank five bundle $\Vt$.  We
must ensure that $\Vt$ remains equivariant under the $\IZ_{12}$ action, such
that it still descends to a bundle $V$ on the quotient space $X$
\cite{Anderson:2009mh,Donagi:2003tb}, and that it remains polystable, so
that we still obtain a solution to the Hermitian-Yang-Mills equations.  Since
deformations cannot change the characteristic classes of a bundle, we
automatically retain the desirable feature of having three net generations of
particles, but the unbroken gauge group is reduced to $SU(5)_\GUT$,
which can then be further broken to exactly $\GSM$ by $\IZ_{12}$-valued
Wilson lines.

There are a number of discrete choices involved in the above, which lead
to many candidate models.  First, note that the bundle
$\cO_{\Xt}{\oplus}\cO_{\Xt}$ is generated by two constant global sections,
so $\IZ_{12}$ can act on it through any of its two-dimensional
representations, corresponding to multiplication of each factor by a twelfth
root of unity.  This gives many choices of equivariant structure on
$\cO_{\Xt}{\oplus}\cO_{\Xt}$, but we will see later that stable deformations will
only exist if the representation is chosen as a sub-representation of that on
$H^{1,1}(\Xt)$, which is a sum of eight distinct one-dimensional
representations.  So there are $\binom{8}{2} = 28$ possibilities.
Another way of saying this is that the space of stable $\IZ_{12}$-equivariant
deformations of $T\Xt{\oplus}\cO_{\Xt}{\oplus}\cO_{\Xt}$ consists of 28 disjoint
branches.  The second choice is the particular $\IZ_{12}$-valued Wilson lines
which break $SU(5)_{\GUT}$ to $\GSM$, for which there are 11 possibilities.
Altogether, then, there are $11\times 28 = 308$ distinct families of equivariant
bundles.

As well as having a stable gauge bundle, a heterotic model must satisfy the
anomaly cancellation condition
\begin{equation*}
    c_2(T\widetilde X) - c_2(\Vt) - c_2(\Vt_{\mathrm{hid.}}) = [C] ~,
\end{equation*}
where $\Vt_{\mathrm{hid.}}$ is another stable bundle in the hidden sector
and $[C]$ is the homology class of some complex curve in $\widetilde X$,
on which M5-branes can be wrapped in the strongly-coupled regime.  In our
case, since $\Vt$ has the same characteristic classes as the tangent bundle,
this is satisfied by a trivial bundle\footnote{A trivial gauge bundle in the hidden
sector leads to a pure $E_8$ super-Yang-Mills sector in four dimensions, which
becomes strongly coupled and breaks supersymmetry at a high scale via
gaugino condensation.  This scale could be lowered by turning on Wilson
lines in the hidden sector, to break $E_8$ to some smaller gauge group.}
 in the hidden sector and $[C] = 0$ i.e. no
5-branes.

Once we have chosen the gauge bundle, we must of course find the
corresponding matter spectrum.  The first step is to calculate the cohomology
group $H^1(\Xt, \Vt)$, corresponding to massless chiral multiplets in the
$\rep{10}$ of $SU(5)_{\GUT}$.  It is also necessary to keep track of the $\IZ_{12}$
action on this space, which combines with the Wilson lines to give the massless
spectrum on the quotient space $X$.  The index theorem guarantees that we
will have three net copies of the $\rep{10}$, but some models will also have
unwanted vector-like pairs of fields originating in this representation.

Finally, we must ask whether these models solve the doublet-triplet splitting
problem.  The cohomology group $H^1(\Xt, \L^2 \Vt)$ gives rise to massless
chiral $\conjrep{5}$'s, and again it is important to keep track of the group
action.  Here, as above, we are guaranteed to have three \emph{net} copies
of this representation, but now we also want an extra vector-like pair of
doublets, to play the role of the Higgs fields of the MSSM.

The remainder of this paper is devoted to carrying out the above procedure
in all its gory detail.  For the impatient reader, we reveal here that
in the end, we find eight bundles which satisfy all the above constraints, and
therefore give models with exactly the massless spectrum of the MSSM.
In fact, these models are possibly unique in this respect among deformations
of standard embedding models; we defer the explanation of this statement
to \sref{sec:conclusion}, as it relies on certain details of the calculations we
present.

\section{The manifold}\label{sec:manifold}

Our compactification manifold $X$ is a quotient by the group $\IZ_{12}$ of a
manifold $\Xt$ which has Hodge numbers $\hodgenos = (8,44)$.  In this section
we will construct $\Xt$ and describe the action of the group.  As we will see,
$\Xt$ is constructed as an anticanonical hypersurface in a toric fourfold; for
detailed general accounts of this method of constructing Calabi-Yau manifolds,
see \cite{Batyrev:1994hm,Kreuzer:2000xy}.  We will therefore make
use of the machinery of toric geometry; reviews for physicists can be
found in \cite{Aspinwall:1993nu,Bouchard:2007ik,Hori:2003ic}, and a number
of good textbooks are also available, including \cite{Fulton,CoxLittleSchenck}.

The manifold $\Xt$ was first constructed as a CICY (complete intersection
Calabi-Yau manifold in a product of projective spaces) in \cite{Candelas:1987kf},
and a free $\IZ_3$ quotient was found in \cite{Candelas:1987du}.  The quotient
group was extended to $\IZ_6$ in \cite{Candelas:2008wb}, and finally to the
two order-twelve groups $\IZ_{12}$ and $\Dic_3 \cong \IZ_3\rtimes\IZ_4$ as part
of the effort to classify all free quotients of CICY's \cite{Braun:2010vc}.  It is the
quotients by these last two groups that give rise to manifolds with $\chi = -6$, and
therefore to three-generation models via the standard embedding.

\begin{figure}[t]
\begin{center}
    \includegraphics[width=.4\textwidth]{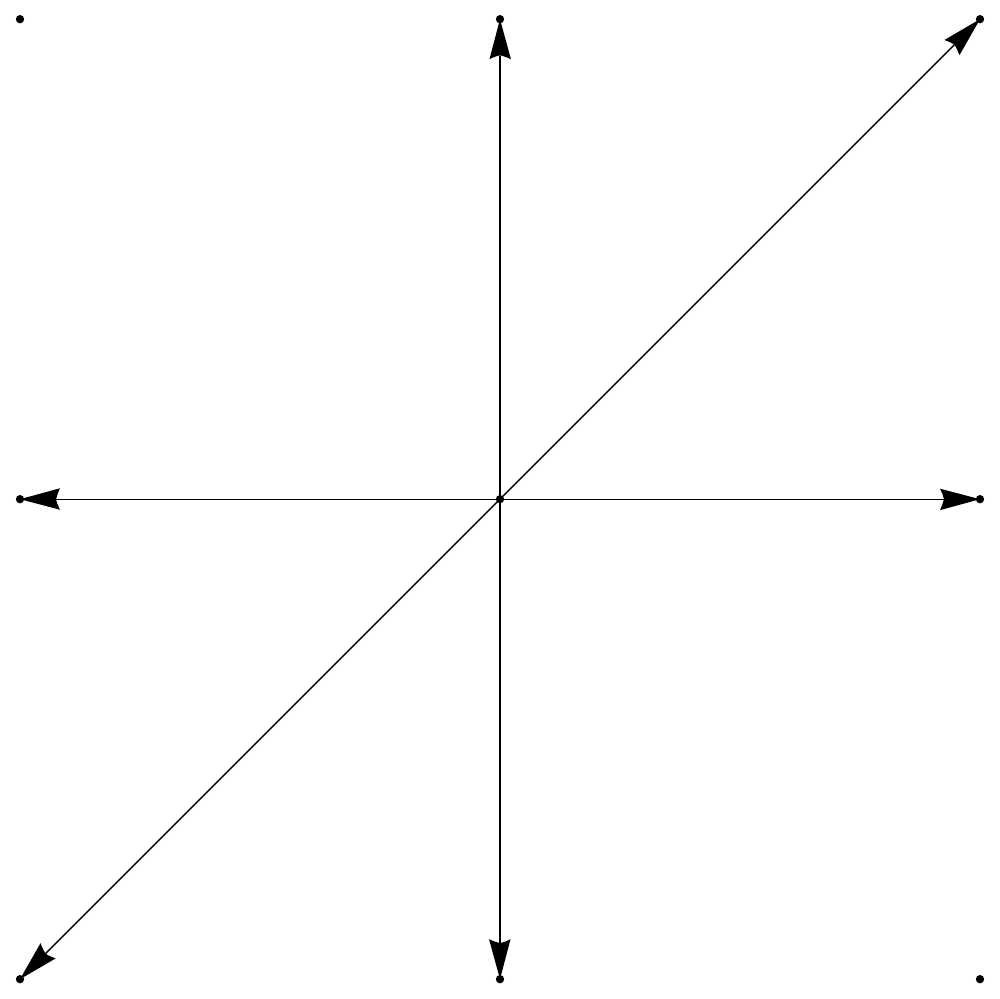}\\
    \place{4.5}{1.47}{$\cD_1$}
    \place{4.5}{2.8}{$\cD_2$}
    \place{3.15}{2.85}{$\cD_3$}
    \place{1.72}{1.5}{$\cD_4$}
    \place{1.83}{.13}{$\cD_5$}
    \place{3.15}{0.1}{$\cD_6$}
    \place{3.55}{1.62}{$\s_1$}
    \place{3.3}{1.8}{$\s_2$}
    \place{2.95}{1.65}{$\s_3$}
    \place{2.75}{1.28}{$\s_4$}
    \place{3.}{1.05}{$\s_5$}
    \place{3.35}{1.25}{$\s_6$}
\parbox{.8\textwidth}{
\caption{\small \label{fig:dP6_fan}
The fan for $\dP_6$, with the rays labelled by the corresponding toric divisors.
The two-dimensional cones $\s_a$ are also labelled.
}}
\end{center}
\end{figure}

Although $\Xt$ can be represented as a CICY, we will find it more useful to
construct it as an anticanonical hypersurface in the toric fourfold
$Z = \dP_6{\times}\dP_6$, where $\dP_6$ is the del Pezzo surface of degree
six.\footnote{The surface referred to here as $\dP_6$ is therefore the complex
projective plane $\IP^2$ blown up at 3 points in general position; in some other
works, including papers by some of the present authors, it is referred to as
$\dP_3$.}  The fan for $\dP_6$ is shown in \fref{fig:dP6_fan}; its one-dimensional
cones are generated by the six vectors
\begin{equation}\label{eq:dP6_vertices}
    \big\{\n_a \big\}_{a=1}^6 =  \{\,(1,0),\,(1,1),\,(0,1),\,(-1,0),\,(-1,-1),\,(0,-1)\,\}~,
\end{equation}
which are the vertices of a hexagon.  We will also have use for the
vertices of the dual hexagon, given by
\begin{equation} \label{eq:dual_hex}
     \big\{ \m_a \big\}_{a=1}^6 = \{\, (0,1),\, (-1,1),\, (-1,0),\, (0,-1),\, (1,-1),\, (1,0) \}~.
\end{equation}
Our toric fourfold $Z$ is a product of two copies of $\dP_6$, so its fan
is just a product of two copies of the fan for $\dP_6$.  Another way
to say this is that it is given by cones over the faces of a four-dimensional
polytope $\nabla$, with vertices
\begin{equation*}
    \mathrm{vert}(\nabla) = \{~ (\n_a, 0)  ~\}_{a=1}^6 \cup \{~ (0, \n_a) ~\}_{a=1}^6 ~.
\end{equation*}
Besides the origin, these twelve vertices are the only lattice points contained
in the polytope $\nabla$.  They correspond to the twelve toric divisors on
$Z$; we will write $\cD_a$ for the divisor corresponding to $(\n_a, 0)$, and
$\widetilde \cD_a$ for the divisor corresponding to $(0, \n_a)$.  It will also be
convenient to have notation encompassing all twelve vertices or divisors at
once, so define
\begin{equation}\label{eq:nabla_vertices}
    v_i = \left\{ \begin{array}{l c l}
        (\n_i, 0) &,& i = 1,\ldots,6 \\
        (0, \n_{i-6}) &,& i = 7,\ldots,12 \end{array}\right.  ~,
\end{equation}
and let $D_i$ be the divisor corresponding to $v_i$.  We also have
corresponding homogeneous coordinates $\{z_i\}_{i=1}^{12}$.

The twelve vectors $v_i$ span a four dimensional space, so there are eight
linear relations between them, which we write as
\begin{equation*}
    \sum_{i=1}^{12} \, Q_{\a i}\, v_i = 0 ~,~ \a = 1,\ldots,8 ~.
\end{equation*}
The matrix $Q$ is only determined up to the left action of $GL(8,\IZ)$; one
possibility is to take it to be, in block form,
\begin{equation} \label{eq:charge_matrix}
    Q_{\a i} ~=~
        \left( \begin{array}{cc}
        Q' & 0 \\
        0 & Q' \end{array} \right) ~,
\end{equation}
\noindent where
\begin{equation*}
    Q' = \left( \begin{array}{rrrrrr}
     1 & 0 & \+0 & \+1 & \+0 & \+0 \\
     0 & 1 & 0 & 0 & 1 & 0 \\
     0 & 0 & 1 & 0 & 0 & 1 \\
     1 & -1 & 1 & 0 & 0 & 0
    \end{array} \right) ~.
\end{equation*}

We can now follow Batyrev's procedure to construct a family of Calabi-Yau
hypersurfaces in $Z$ \cite{Batyrev:1994hm}.  Let $M$ be the dual lattice to
$N$, and denote by $M_\IR$ and $N_\IR$ the real vector spaces in which
the lattices lie.  Then the dual polytope to $\nabla$ is defined in $M_\IR$
by
\begin{equation*}
    \D = \{ u \in M_\IR ~\vert~ \langle u, v \rangle \geq -1 ~\forall~ v \in \nabla \} ~.
\end{equation*}
For the present example, this is simple to describe explicitly.  Note that the dual
hexagon we mentioned earlier, with vertices given by \eqref{eq:dual_hex},
contains seven lattice points including the origin.  The polytope $\D$ therefore
contains 49 lattice points, given by pairs of these.  Of these points, 36 are the
vertices, given by
\begin{equation*}
    \mathrm{vert}(\D) = \{~ (\m_a, \m_b)  ~\}_{a,b=1}^6 ~.
\end{equation*}
The thirteen additional lattice points consist of the origin, and points of the
form $(\mu_a, 0)$ or $(0, \mu_a)$; the 12 three-faces of $\D$ are hexagonal
prisms each with two hexagonal faces and six rectangular faces, and these extra
points arise as one point interior to each of the hexagonal two-faces
\cite{Braun:2009qy}.  Altogether, the 49 lattice points of $\D$ correspond to a
basis of the space of global sections of the anticanonical bundle
$\o_Z^{-1} \cong \cO_Z(\sum_i D_i)$, via the prescription
\begin{equation} \label{eq:monomials}
    u ~\to~ \prod_i \, z_i^{\langle u, v_i \rangle + 1} ~.
\end{equation}

\vspace{-3ex}
\subsection{The group action}

The group $\IZ_{12}$, being cyclic, has a very simple representation
ring.  All of its irreducible representations are one-dimensional, and there
are twelve altogether, corresponding to the twelfth roots of unity.  We will
denote by $\rep{n}$ the representation which sends the generator $g_{12}$
of $\IZ_{12}$ to $e^{\frac{\pi\ii}{6} n}$, where $n$ takes the values $0$ through
$11$.

The action of $\IZ_{12}$ on the lattice $N$ (in which $\nabla$ lives) is
generated by the matrix
\begin{equation*}
    A_N = 
        \left( \begin{array}{r r r r}
            0 & 0 & 0 & 1 \\
            0 & 0 & -1 & 1 \\
            0 & -1 & 0 & 0 \\
            1 & -1 & 0 & 0
        \end{array} \right) ~.
\end{equation*}
Geometrically, this exchanges the two hexagons, then rotates the first
clockwise by one-sixth of a turn, and the second anti-clockwise by one
third of a turn.  The $\IZ_{12}$ action is therefore transitive on the set of
toric divisors, being given by
\vskip-2ex
\begin{equation*}
    g_{12} ~:~ \cD_a \to \widetilde \cD_{a+2} ~,~ \widetilde \cD_a \to \cD_{a-1} ~,
\end{equation*}
where the subscripts are understood modulo 6.  If we instead use the
notation $D_i$, $i=1,\ldots, 12$ for the divisors, as in
\eqref{eq:nabla_vertices}, the group action is explicitly given by the
permutation
\begin{equation} \label{eq:permutation}
    D_1 \to D_9 \to D_2 \to D_{10} \to D_3 \to D_{11} \to D_4 \to D_{12} \to D_5
         \to D_7 \to D_6 \to D_8 \to D_1 ~,
\end{equation}
with the homogeneous coordinates being permuted in the same way.

To find the invariant anticanonical hypersurfaces, we need to know how
the group acts on the dual lattice $M$, and hence on the monomials
\eqref{eq:monomials}.  This is determined by demanding that the duality
pairing is preserved, i.e.
\begin{equation*}
    \langle A_M u, A_N v\rangle ~=~ \langle u, v\rangle ~~\forall~~
        u \in M ~,~ v \in N ~.
\end{equation*}
In matrix terms, this implies $A_M = \big(A_N^{-1}\big)^T$, which is
explicitly
\begin{equation*}
    A_M =
        \left(\hspace{-.5em} \begin{array}{rrrr}
            0 & 0 & 1 & 1 \\
            0 & 0 & -1 & 0 \\
            -1 & -1 & 0 & 0 \\
            1 & 0 & 0 & 0 \end{array} \right) ~.
\end{equation*}
The 49 lattice points in $\D$ fall into five orbits under the action of
$\IZ_{12}$.  Obviously the origin is fixed by the group, and therefore
corresponds to an invariant monomial $f_0$, while the other 48 points
fall into four orbits of length twelve.  Each orbit yields a unique invariant
polynomial, given by the sum of the twelve monomials which get
permuted.  This yields four more invariant polynomials
$f_1, \ldots , f_4$, with the most general invariant polynomial being
given by $f = \sum_{\l=0}^4 a_\l f_\l$.  

It can be checked (see e.g. \cite{Braun:2009qy}) that for a generic
choice of the coefficients $a_\l$, the corresponding hypersurface $\Xt$ is
smooth.  These coefficients are therefore projective coordinates on
the four-dimensional moduli space of a family of smooth quotient
manifolds $X$.  For the remainder of the paper, we will choose them to
be generic integers; none of our results depend on this choice.

We can also now calculate the group action on
$H^{1,1}(\Xt) \cong H^1(\Xt, \O^1\Xt)$, which will be important later.  First
we note that $H^{1,1}(\Xt) \cong H^{1,1}(Z)$, which follows from the
Lefschetz hyperplane theorem after noting that $X$ is an anticanonical
hypersurface in $Z$, and that the anticanonical bundle of $Z$ is very
ample.  We then have the general result for smooth compact toric
varieties,
\begin{equation*}
    H^{1,1}(Z) ~\cong~ \frac{\bigoplus_i \IC\cdot v_i}{M\otimes_{\IZ}\IC} ~.
\end{equation*}
The permutation action of $\IZ_{12}$ on the toric divisors corresponds to the
regular representation $\mathrm{Reg}_{\IZ_{12}} = \bigoplus_n \rep{n}$, and
by diagonalising the matrix $A_M$, one sees that $\IZ_{12}$ acts on
$M{\otimes}_{\IZ}\IC$ via the faithful representation\footnote{This also
follows from noting that $A_M$ is real, and neither $A_M^4$ nor $A_M^6$
has $1$ as an eigenvalue.}
$\rep{1}\oplus\rep{5}\oplus\rep{7}\oplus\rep{11}$.  We conclude that
$H^{1,1}(Z)$, and hence $H^{1,1}(\Xt)$, transforms under the difference
of these representations:
\begin{equation}\label{eq:1,1_rep}
    H^{1,1}(\Xt,\IC) ~\cong~ H^{1,1}(Z,\IC) ~\sim~
        \rep{0}\oplus\rep{2}\oplus\rep{3}\oplus\rep{4}\oplus\rep{6}\oplus\rep{8}\oplus\rep{9}\oplus\rep{10} ~.
\end{equation}
We will rederive this result later by a different method.

\newpage
\section{Deforming the gauge bundle}\label{sec:bundle}

In this section we will explicitly construct our desired family of rank-five
bundles on the covering space $\Xt$, and argue that they are stable.
The reader wishing to avoid these details may skip to the commutative
diagram in \eqref{eq:V_diagram}, which provides the ultimate definition
of the bundles, and serves as the basis for the cohomology calculations
of \sref{sec:spectrum}.

\subsection{Stability and deformations}

Perhaps the most difficult part of constructing appropriate vector bundles
for heterotic string theory is ensuring that the bundles are slope stable, and
therefore, by the Donaldson-Uhlenbeck-Yau theorem, admit a
Hermitian-Yang-Mills connection \cite{Donaldson:1985zz,UhlenbeckYau}.
Here we are interested in bundles which are
deformations of $T\Xt\oplus\cO_{\Xt}^{\oplus r}$, where $\Xt$ is a Calabi-Yau
manifold, and in this case Li and Yau found much simpler necessary
and sufficient conditions for a solution \cite{Li:2004hx}.  Before turning to
the details of our bundles, we will briefly review their result.

Consider a one-parameter family of bundles $\Vt_s$, such that
$\Vt_0 = T\Xt{\oplus}\cO_{\Xt}^{\oplus r}$.  Then the `tangent' to this family of
bundles at $s=0$ is given by the Kodaira-Spencer deformation class,
which lives in $H^1(\Xt, \Vt_0\otimes\dual{\Vt_0\!})$.  Using the decomposition
of $\Vt$, and the fact that $H^1(\Xt, \cO_{\Xt}) = 0$ since $\Xt$ is Calabi-Yau, this
becomes
\begin{equation}\label{eq:Kodaira-Spencer}
    H^1(\Xt, \Vt_0\otimes\dual{\Vt_0\!}) = H^1(\Xt,\O^1\Xt{\otimes}T\Xt)\oplus
        H^1(\Xt,T\Xt)^{\oplus r}\oplus H^1(\Xt, \O^1\Xt)^{\oplus r} ~.
\end{equation}
The Li-Yau conditions are that the projections of the Kodaira-Spencer class
onto each of the last two terms consist of $r$ linearly independent
elements of the respective cohomology groups.

To get some intuition for what this means, note that $H^1(\Xt, T\Xt)$ is
naturally isomorphic to $\Ext^1(\cO_{\Xt}, T\Xt)$, which parametrises extensions
of $\cO_{\Xt}$ by $T\Xt$, i.e. bundles $B$ which fit into a short exact sequence
\begin{equation*}
    0 \longrightarrow T\Xt \longrightarrow B \longrightarrow \cO_{\Xt} \longrightarrow 0 ~.
\end{equation*}
If $B \cong T\Xt{\oplus}\cO_{\Xt}$, the extension is called \emph{split}, and this
corresponds to the zero element of $\Ext^1(\cO_{\Xt}, T\Xt)$.  If we replace
$\cO_{\Xt}$ by $r \cO_{\Xt} := \cO_{\Xt}^{\oplus r}$, then we get $r$ copies of the
same $\Ext^1$ group.  An element of this group is therefore an $r$-tuple of
elements of $\Ext^1(\cO_{\Xt}, T\Xt)$, and these $r$ elements are linearly
independent if and only if the corresponding extension is \emph{completely
non-split}, i.e. cannot be written as $B \cong B'\oplus\cO_{\Xt}$ for some
bundle $B'$.  Analogous comments apply to $H^1(\Xt, \O^1\Xt)$, which
parametrises `opposite' extensions i.e. extensions of $T\Xt$ by $\cO_{\Xt}$.

An extension of $T\Xt$ by $\cO_{\Xt}$ or vice versa can never be stable, but
if the Kodaira-Spencer class of the family $\Vt_s$ has a non-zero piece in each
of $H^1(\Xt,T\Xt)^{\oplus r}$ and $H^1(\Xt, \O^1\Xt)^{\oplus r} $, then
$\Vt_s$ for $s\neq 0$ has no such simple interpretation, and may be stable.
The theorem of Li and Yau asserts that $\Vt_s$ is stable for small enough
$s$ precisely if the components of each of these $r$-tuples are linearly
independent.

\subsection{Constructing the deformations}\label{sec:deformations}

As we wish to ultimately construct a rank-five bundle on $\Xt$ which is a deformation
of $T\Xt{\oplus}\cO_{\Xt}{\oplus}\cO_{\Xt}$, it is necessary to first give an explicit
description of the tangent bundle itself.  We begin with the normal bundle short
exact sequence, noting that the normal bundle to $\Xt$ inside $Z$ is isomorphic to
the restriction of the anticanonical bundle of $Z$,
\begin{equation}\label{eq:normal_bundle}
    0 \longrightarrow T\Xt \longrightarrow TZ\rest{\Xt} \stackrel{df}{\longrightarrow}
        \cO_{\Xt}\big(\sum_i \, D_i\big) \longrightarrow 0 ~.
\end{equation}
In turn, the bundle $TZ$ is simply described by a generalisation of the Euler
sequence, which applies to all smooth toric varieties:
\begin{equation}\label{eq:toric_euler}
    0 \longrightarrow 8 \cO_{Z} \longrightarrow \bigoplus_{i=1}^{12} \, \cO_{Z}(D_i) \longrightarrow
        TZ \longrightarrow 0 ~,
\end{equation}
where the first map here is defined on global sections by a matrix with entries
$Q_{\a,i} z_i$ (no summation on $i$), where $Q_{\a,i}$ is the toric charge
matrix from \eqref{eq:charge_matrix}.  We will pause here briefly to
explain this in a little more detail.

Following Cox's prescription \cite{Cox:1993fz}, the space $Z$ can be constructed
from the affine space $\IC^{12}$ parametrised by $\{z_i\}$ by deleting certain
coordinate planes, and then imposing the equivalence relations
\begin{equation} \label{eq:scalings}
    (z_1, \ldots,z_{12}) \sim (\l^{Q_{\a,1}}z_1,\ldots,\l^{Q_{\a,12}}z_{12}) ~\forall~ \l \in \IC^*
        ~,~\a = 1,\ldots, 8 ~.
\end{equation}
Vector fields on $\IC^{12}$ are of the form
$\sum_i \, h_i(z) \frac{\partial}{\partial z_i}$, but these will only descend to $Z$
if they are invariant under \eqref{eq:scalings}, i.e. if $h_i(z)$ transforms like
$z_i$ under all the rescalings, which is equivalent to saying that $h_i(z)$ is
a section of $\cO_Z(D_i)$.  So vector fields on $Z$ are given by sections of
$\bigoplus_i \cO_Z(D_i)$, but some of these may correspond to the zero vector
field.  This will be the case when the vector field on $\IC^{12}$ is tangent to the
equivalence classes defined by \eqref{eq:scalings}, i.e. when
$h_i = k\, Q_{\a, i} z_i$ for some constant $k$.  This argument can be made
rigorous, and the result is that the sequence in \eqref{eq:toric_euler} is exact.

So we have an eight-dimensional space of `Euler vectors' spanned by
\begin{equation}\label{eq:euler_vectors}
    \left\{ E_\a = \sum_{i=1}^{12}\, Q_{i\a} z_i \frac{\partial}{\partial z_i} \right\}_{\a = 1}^8 ~,
\end{equation}
which corresponds to the space of global sections of $8\cO_{\Xt}$ in
\eqref{eq:toric_euler}.  These will be very important as we proceed.

We will construct our desired bundles by adapting the techniques of
\cite{Li:2004hx,Donagi:2006yf}.  First, notice that after restricting
the Euler sequence \eqref{eq:toric_euler} to the hypersurface $\Xt$, we can
compose the last map in this sequence with the differential $df$ from
\eqref{eq:normal_bundle} to get a map
\begin{equation*}
    \Phi_0 : \bigoplus_i  \cO_{\Xt}(D_i) \to \cO_{\Xt}(\sum_i D_i) ~.
\end{equation*}
Then if we denote by $\Ft_0$ the bundle which is the kernel of this
map,\footnote{The kernel of a map between vector bundles may fail to be a
bundle itself, but instead be a more general coherent sheaf.  This does not
occur here or in similar situations later in the paper, so we can ignore this
subtlety.} a simple diagram-chasing argument shows that the dashed arrows
in the following commutative diagram can be filled in such that all the rows
and columns are exact
\begin{equation}\label{eq:TX_diagram}
   \begin{diagram}[heads=littleblack]
         &  	  &  0			     &			       &  0			&			&     			&		&  \\
         &  	  &  \dTo		     &			       &  \dTo			&			&     			&		&  \\
         &  	  &  8\,\cO_{\Xt} &  \rTo^{=} 	       &  8\,\cO_{\Xt}&			&     			&		&  \\
         &  	  &  \dDashto	     &			       &  \dTo			&			&     			&		&  \\
      0 &  \rTo  &  \Ft_0		     &  \rTo		       & \bigoplus_{i=1}^{12} \cO_{\Xt}(D_i)
      			&  \rTo^{\Phi_0}	&  \cO_{\Xt}\left(\sum_{i=1}^{12} D_i\right)	&  \rTo 	&  0 \\
         &	  &  \dDashto	     &			       &  \dTo			&			&  \dTo>\Vert        &		&  \\
      0 &  \rTo  &  T\Xt		     &  \rTo		       &  TZ\vert_{\Xt}		&  \rTo^{df}		&  \cO_{\Xt}\big(\sum_{i=1}^{12} D_i\big)	&  \rTo	&  0  \\
         &	  &  \dTo		     &			       &  \dTo			&			&			&		&  \\
         &	  &  0			     &			       &  0			&			&			&		&  \\
   \end{diagram} \raisebox{-26ex}{~~.}
\end{equation}
Note that the exactness of rows and columns here implies that
$8\cO_{\Xt} \subset \ker\Phi_0$, which says that $\Phi_0$ annihilates
all eight Euler vectors in \eqref{eq:euler_vectors}.  This is a result of
the generalised Euler identities $E_\a(f) \propto f$, and the fact that
all bundles are restricted to $\Xt$, where $f \equiv 0$.

Exactness of the leftmost column of \eqref{eq:TX_diagram} means
that $\Ft_0$ is an extension of $T\Xt$ by the rank-eight trivial bundle
$8\cO_{\Xt}$, and it is not too hard to argue that it is a completely
non-split extension.  Indeed, suppose that it split, so that
$\Ft_0 \cong \Ft'_0\oplus\cO_{\Xt}$.  Then the projection onto the second
summand would correspond to a non-trivial element of
$\Hom(\Ft_0, \cO_{\Xt}) \cong H^0(\Xt, \dual{\Ft_0\!}\otimes\cO_{\Xt})
\cong H^0(\Xt,\dual{\Ft_0\!})$, but the calculations of
\aref{app:X_cohomology} show that this last group is trivial.  We
conclude that no such split can exist.

The bundle $\Ft_0$ we have just defined has vanishing first Chern
class, but has rank eleven and, being an extension by $8\cO_{\Xt}$,
is unstable.  Let us solve the first of these problems, and obtain a rank
five bundle instead.  We need only select a trivial rank-six
sub-bundle\footnote{We will be more specific about the choice of
$6\cO_{\Xt}$ inside $8\cO_{\Xt}$ in \sref{sec:equivariance} when we
discuss group equivariance of the various bundles here.}
$6\cO_{\Xt} \subset 8\cO_{\Xt} \subset \Ft_0$, and define
$\Vt_e = \Ft_0/6\cO_{\Xt}$.  Then $\Vt_e$ is a completely non-split
extension\footnote{If the extension split, then, as above, it would imply
a non-trivial morphism $\Vt_e \to \cO_{\Xt}$, but composing this with
the surjection $\Ft_0 \to \Vt_e$ would give a non-trivial element of
$H^0(\Xt, \dual{\Ft_0\!})$, which we have already stated does not
exist.} of $T\Xt$ by $2\cO_{\Xt}$.

The bundle $\Vt_e$ can be considered to be a deformation of
$T\Xt{\oplus}\cO_{\Xt}{\oplus}\cO_{\Xt}$, and the argument above shows
that its Kodaira-Spencer class is a non-degenerate element of
$H^1(\Xt, \O^1\Xt)^{\oplus 2}$, thereby satisfying half of the Li-Yau
conditions.  We now need to further deform $\Vt_e$ so that the part of
the Kodaira-Spencer class living in $H^1(\Xt, T\Xt)^{\oplus 2}$ is also
non-degenerate.  We can do this by deforming the map $\Phi_0$,
which is defined by the first derivatives of the polynomial $f$, to a more
general morphism
\begin{equation*}
    \Phi ~:~ \bigoplus_{i=1}^{12} \cO_{\Xt}(D_i) \to \cO_{\Xt}\big(\sum_{i=1}^{12} D_i\big) ~,
\end{equation*}
and defining $\Ft = \ker \Phi$.  If we demand that $\Phi$ still annihilates
our chosen sub-bundle $6\cO_{\Xt} \subset 8\cO_{\Xt}$, then we can still
define a rank-five bundle $\Vt = \Ft/6\cO_{\Xt}$; this can be expressed by
the following short exact sequence,
\begin{equation} \label{eq:seq_F}
    0 \longrightarrow 6 \cO_{\Xt} \longrightarrow \Ft \longrightarrow \Vt \longrightarrow 0 ~.
\end{equation}
The general $\Vt$ defined this way is a deformation of $\Vt_e$, and therefore
of $\Vt_0 = T\Xt{\oplus}\cO_{\Xt}{\oplus}\cO_{\Xt}$.  Note that the map $\Phi$
is an element of
\begin{align}
    &\Hom\Big(\bigoplus_{i=1}^{12} \cO_{\Xt}(D_i)~,~ \cO_{\Xt}\big(\sum_{i=1}^{12} D_i\big)\Big)
        ~\cong~ H^0\Big(\Xt, ~\bigoplus_{j=1}^{12} \cO_{\Xt}\big(\sum_{i=1}^{12} D_i - D_j\big)\Big)\notag \\[1ex]
    \cong&~~~ \bigoplus_{j=1}^{12}H^0\Big(\Xt, ~\cO_{\Xt}\big(\sum_{i=1}^{12} D_i - D_j\big)\Big) ~~\cong~
         \IC^{12\times 35} ~=~ \IC^{420} ~,  \label{eq:Phi_space}
\end{align}
where the dimension follows from results of \aref{app:cohomology}.  Demanding
that $\Phi$ annihilate six specific global sections restricts it to a 132-dimensional
subspace.

We must now ask whether the new $\Phi$ is sufficient to satisfy the other
half of the Li-Yau conditions.  This could be demonstrated along the lines of
the original paper \cite{Li:2004hx}, by constructing a smooth moduli space
which interpolates between non-split extensions of $T\Xt$ by $2\cO_{\Xt}$ and
vice versa, and contains our more general bundles $\Vt$.  Here though, we will
simply note that any stable bundle with vanishing first Chern class has no global
sections, and this condition is also sufficient in this case.  We give an alternative
argument for this, and a simpler proof of stability, in \aref{app:stability}.

So we need only arrange that $H^0(\Xt, \Vt) = 0$.  Consider the long exact sequence
in cohomology which follows from \eqref{eq:seq_F}.  Since $\Xt$ is Calabi-Yau, we
have $H^1(\Xt, \cO_{\Xt}) = 0$, so that the first few terms are
\begin{equation*}
    0 \longrightarrow \IC^6 \longrightarrow H^0(\Xt, \Ft) \longrightarrow H^0(\Xt, \Vt)
        \longrightarrow 0 \longrightarrow \ldots ~~~.
\end{equation*}
To obtain $H^0(\Xt, \Vt) = 0$, we therefore need $H^0(\Xt, \Ft) \cong \IC^6$.
In words, we must demand that, when applied to global sections of
$\bigoplus_{i=1}^{12} \cO_{\Xt}(D_i) $, $\Phi$ annihilates \emph{precisely}
our chosen six-dimensional subspace.  If we make a specific choice of the
parameters defining $\Phi$, this can be checked by computer, and in all
examples, a general enough $\Phi$ exists such that it is true.  This ensures
that $\Vt$ is stable.

There is a different, equivalent way to define $\Vt$, which will prove
useful for later calculations.  We start with \eqref{eq:toric_euler}, which
defines the tangent bundle of the ambient space, $TZ$, as a quotient
$\oplus_{i=1}^{12} \cO_Z(D_i)/8\cO_Z$, and instead define a rank-six
bundle $\Gt$ by dividing only by our chosen sub-bundle $6\cO_Z$.
This is captured by a short exact sequence,
\begin{equation}\label{eq:seq_1}
    0 \longrightarrow 6\cO_Z \longrightarrow \bigoplus_{i=1}^{12} \cO_Z(D_i)
        \longrightarrow \Gt \longrightarrow 0 ~,
\end{equation}
which we can also restrict to $\Xt$.  Since we demanded that
$\ker \Phi \supset 6\cO_{\Xt}$, $\Phi$ induces a well-defined map from
$\Gt\rest{\Xt}$ to $\cO_{\Xt}(\sum_{i=1}^{12} D_i)$, and
$\Vt$ is precisely the kernel of this map, giving another short exact
sequence,
\begin{equation}\label{eq:seq_2}
    0 \longrightarrow \Vt \longrightarrow \Gt\rest{\Xt}
        \longrightarrow \cO_{\Xt}(\sum_{i=1}^{12} D_i) \longrightarrow 0 ~.
\end{equation}

We also note in passing that $\Vt$ can in fact be defined in one step as the
cohomology of the (non-exact) sequence
\begin{equation*}
    6\cO_{\Xt} \longrightarrow \bigoplus_{i=1}^{12} \cO_{\Xt}(D_i) \longrightarrow \cO_{\Xt}(\sum_{i=1}^{12} D_i) ~.
\end{equation*}
This definition of $\Vt$ would allow a worldsheet description of these
compactifications via the linear sigma model \cite{Distler:1995mi}.

The two short exact sequences in \eqref{eq:seq_1} and \eqref{eq:seq_2}
can be intertwined with those defining $\Vt$ in terms of $\Ft$, to yield a
commutative diagram analogous to that in \eqref{eq:TX_diagram},
\begin{equation}\label{eq:V_diagram}
   \begin{diagram}[heads=littleblack]
         &  	  &  0			     &			       &  0			&			&     			&		&  \\
         &  	  &  \dTo		     &			       &  \dTo			&			&     			&		&  \\
         &  	  &  6\,\cO_{\Xt} &  \rTo^{=} 	       &  6\,\cO_{\Xt}&			&     			&		&  \\
         &  	  &  \dTo		     &			       &  \dTo			&			&     			&		&  \\
      0 &  \rTo  &  \Ft		     &  \rTo		       & \bigoplus_{i=1}^{12} \cO_{\Xt}(D_i) 		&  \rTo^{\Phi}
      			&  \cO_{\Xt}\big(\sum_{i=1}^{12} D_i\big)	&  \rTo 	&  0 \\
         &	  &  \dTo		     &			       &  \dTo			&			&  \dTo>\Vert        &		&  \\
      0 &  \rTo  &  \Vt		     &  \rTo		       &  \Gt\rest{\Xt}		&  \rTo
      			&  \cO_{\Xt}\big(\sum_{i=1}^{12} D_i\big)	&  \rTo	&  0  \\
         &	  &  \dTo		     &			       &  \dTo			&			&			&		&  \\
         &	  &  0			     &			       &  0			&			&			&		&  \\
   \end{diagram} \raisebox{-26ex}{~~.}
\end{equation}
When interpreting this diagram, one should consider $\Ft$ and
$\Gt$ to be defined by exactness of the middle row and column respectively.
The bundle $\Vt$ can then be defined by the exactness of either the leftmost
column or the bottom row.

\subsection{Bundle equivariance} \label{sec:equivariance}

In this section we will discuss how to make the preceding construction
compatible with the action of the quotient group.  In this context, the word
`equivariant' occurs frequently, and is used to mean two different things.
Firstly, a bundle $\pi : B \to \Xt$ is said to be equivariant under $\IZ_{12}$
if the group generator $g_{12}$ can be made to act on $B$ via a map
$\Psi : B \to B$, such that the following diagram commutes,
\begin{equation*}
    \begin{diagram}[heads=littleblack]
        B		& \rTo^\Psi		& B \\
        \dTo<\pi	& 				& \dTo>\pi \\
        \Xt		& \rTo^{g_{12}}		& \Xt \\
    \end{diagram} \raisebox{-6ex}{~~~,}
\end{equation*}
\vskip1.5ex
\noindent and $\Psi^{12}$ is the identity map on $B$.  The second use of the word
equivariant is to describe a map between two equivariant bundles.  Let
$B_1$ and $B_2$ be two bundles, each equivariant under $\IZ_{12}$,
with the equivariant structures being given by $\Psi_1$ and $\Psi_2$
respectively.  Then a bundle morphism $\Xi : B_1 \to B_2$ is said to be
$\IZ_{12}$-equivariant if $\Xi\circ\Psi_1 = \Psi_2\circ\Xi$.  One might
also say that $\Xi$ `commutes with the group action'.

So far we have constructed a rank-five bundle $\Vt$ on the manifold $\Xt$.
If we want this to correspond to a bundle on the quotient space $X$, we
must ensure that it is equivariant under the action of the quotient group
$\IZ_{12}$.  The easiest way to guarantee this is to ensure that all the
other bundles appearing in \eqref{eq:V_diagram} are equivariant, and that
the maps between them commute with the group action.  Indeed, it is not
hard to convince oneself that if two bundles $B_1$ and $B_2$ are equivariant
under the action of some group, and a morphism $\Xi : B_1 \to B_2$
commutes with the group action, then the kernel and cokernel of $\Xi$
are also equivariant \cite{Anderson:2009mh}.

Begin with the bundle $\bigoplus_{i=1}^{12} \cO_{\Xt}(D_i)$, which is
central to the whole construction.  Since $\IZ_{12}$ permutes the twelve
toric divisors, it naturally acts on this bundle by permuting the twelve direct
summands in the same way.  The bundle $8\cO_{\Xt}$ is embedded in this
rank-twelve bundle via the eight Euler vectors in \eqref{eq:euler_vectors}.
These eight sections generate $8\cO_{\Xt}$, and one can check that the
eight-dimensional space they span is mapped to itself by the $\IZ_{12}$
action, thus defining an action on $8\cO_{\Xt}$ in such a way that the map
$8\cO_{\Xt} \to \bigoplus_{i=1}^{12} \cO_{\Xt}(D_i)$ is automatically
equivariant.  By diagonalising the $8{\times}8$ matrix which acts on the
space of Euler vectors, we find that in terms of $\IZ_{12}$ representations,
\begin{equation}\label{eq:H08O_rep}
    H^0(\Xt, 8\cO_{\Xt}) ~\sim~
        \rep{0}\oplus\rep{2}\oplus\rep{3}\oplus\rep{4}\oplus\rep{6}\oplus\rep{8}\oplus\rep{9}\oplus\rep{10} ~.
\end{equation}
We note in passing that using this result in the long exact cohomology
sequence following from the leftmost column of \eqref{eq:TX_diagram},
we can quickly rederive \eqref{eq:1,1_rep}.  In words, such a calculation
shows that $H^{1,1}(\Xt)$ transforms under the same real representation
as the space of Euler vectors.

The steps above are actually fairly trivial; we have simply defined the
action of $\IZ_{12}$ on $8\cO_{\Xt}$ via its embedding in
$\bigoplus_{i=1}^{12} \cO_{\Xt}(D_i)$, which immediately makes the
embedding map equivariant.  The more involved step is ensuring that
the map $\Phi$ is equivariant.

First note that $\IZ_{12}$ acts on the line bundle $\cO_{\Xt}(\sum_i D_i)$
simply by the permutation of homogeneous coordinates given by
\eqref{eq:permutation}.  This permutation acts on the entire
homogeneous coordinate ring, and we will denote it by
\begin{equation*}
    \s : \IC[z_1, \ldots,z_{12}] \to \IC[z_1, \ldots,z_{12}] ~.
\end{equation*}
If we want to explicitly represent the group action on, say, the global
sections of $\bigoplus_{i=1}^{12} \cO_{\Xt}(D_i)$, we need to apply $\s$
as well as permute the components.  For example, letting $P$ be the
matrix which permutes the components, we have
\begin{equation*}
    g_{12}\big((z_1,0,\ldots, 0)^T\big) ~=~ \s\big(P\cdot(z_1,0,\ldots, 0)^T\big)
        ~=~ (0,\underbrace{\ldots}_6,0,z_9,0,0,0)^T ~.
\end{equation*}
Had we not applied $\s$ here, we would have obtained something
nonsensical, since $z_1$ is not a section of $\cO_{\Xt}(D_9)$.

So if we represent $\Phi$ by a $12{\times}1$ matrix, the condition for
equivariance is that for any section $\t$ of
$\bigoplus_{i=1}^{12} \cO_{\Xt}(D_i)$, represented as a column vector
as above, we have
\begin{equation*}
    \Phi\cdot\big(P\cdot \s(\t)\big) = \s\big(\Phi\cdot \t\big) ~.
\end{equation*}
This may look a little technical, but it simply means, for example, that
$\Phi_1 = \s\big(\Phi_9\big)$, so the twelve components of $\Phi$ are
all related by permuting the coordinates.  Before imposing equivariance,
each component had $35$ parameters, so we find a $35$-dimensional
space of equivariant maps $\Phi$.

Recall that the original map $\Phi_0$, induced by $df$, annihilates all eight
Euler vectors, and that we want $\Phi$ to still annihilate a six-dimensional
subspace.  It is clear that, due to equivariance, this must correspond to
a six-dimensional sub-representation of \eqref{eq:H08O_rep}, of which
there are 28, corresponding to the choice of two of the eight distinct
charges.  It is easy enough to check that in every case, there is an
$11$-dimensional family\footnote{There are three common parameters,
corresponding to equivariant maps which annihilate all eight Euler vectors.
These represent a three-dimensional family of rank-three deformations
i.e. deformations of the tangent bundle of the quotient space $X$.} of
equivariant maps $\Phi$ which annihilate \emph{exactly} the six chosen
Euler vectors, thus guaranteeing stability of the corresponding bundle
$\Vt$, as discussed at the end of \sref{sec:deformations}.  Therefore
there are 28 distinct branches of the moduli space of stable
$\IZ_{12}$-equivariant deformations of $T\Xt\oplus\cO_{\Xt}\oplus\cO_{\Xt}$.

\subsection{Equivariant structures and Wilson lines} \label{sec:Wilson_1}

The various branches of the moduli space of $\Vt$, described previously,
correspond to different choices of equivariant structure on
$T\Xt{\oplus}\cO_{\Xt}{\oplus}\cO_{\Xt}$.  A non-trivial equivariant structure
on $\cO_{\Xt}$, corresponding to some $\IZ_{12}$ representation $\rep{n}$,
means that on traversing a non-contractible path on the quotient space $X$,
we identify sections of $\cO_X$ which differ by the phase
$e^{\frac{\pi\ii}{6}n}$.  So in fact we have a non-trivial flat line bundle $\cL_n$.
Since the holonomy group is identified in physics with the gauge group, this
is precisely what is meant by turning on discrete Wilson lines.

We now come across a slight complication which we have been ignoring
until now.  Take the bundle $TX{\oplus}\cL_{n_1}{\oplus}\cL_{n_2}$, and
ask about the holonomy around a non-contractible path on $X$.  It will
be given by a $5{\times}5$ matrix
\begin{equation*}
    U_5 = \diag(U_3, e^{\frac{\pi\ii}{6}n_1}, e^{\frac{\pi\ii}{6}n_2}) ~,
\end{equation*}
where $U_3 \in SU(3)$ is the holonomy in $TX$.  Although $U_5$ is unitary,
it will not typically lie in $SU(5)$, since $\det U_5 = e^{\frac{\pi\ii}{6}(n_1+n_2)}$.
So we must multiply it by $e^{\frac{\pi\ii}{6}\tilde n}$, where
\begin{equation}\label{eq:ntilde}
    5\tilde n + n_1 + n_2 = 0 \quad \mathrm{mod}~12
        ~~\Rightarrow~ \tilde n = 7(n_1 + n_2) \quad \mathrm{mod}~12~.
\end{equation}
This corresponds to choosing an extra overall phase in the equivariant
structure on $\Vt$, such that it descends to a deformation of the bundle
$\cL_{\tilde n} \otimes (TX{\oplus}\cL_{n_1}{\oplus}\cL_{n_2})$.  This
ensures that the holonomy group is indeed $SU(5)$.

\section{Cohomology and finding the MSSM spectrum}\label{sec:spectrum}

We have constructed various families of stable $\IZ_{12}$-equivariant
bundles, a generic example of which we are denoting by $\Vt$, with
structure group $SU(5)$.  Embedding this $SU(5)$ in $E_8$ gives us a
heterotic model on $\Xt$ with unbroken gauge group $SU(5)_\GUT$.
The decomposition of the adjoint of $E_8$ is
\begin{equation}\label{eq:248}
   \begin{split}
      E_8&~\supset~ \frac{SU(5){\times}SU(5)_\GUT}{\IZ_5} \\[1ex]
      \rep{248}& ~=~(\rep{1},\rep{24})\oplus(\rep{24},\rep{1})
      \oplus(\rep{5},\rep{10})\oplus(\rep{10},\conjrep{5})
      \oplus(\conjrep{5},\conjrep{10})\oplus(\conjrep{10},\rep{5}) ~.
   \end{split}
\end{equation}
The left-handed standard model fermions appear in the
$\rep{10}\oplus\conjrep{5}$ representation of $SU(5)_\GUT$, and
therefore correspond to the fundamental $\rep{5}$ and the rank-two
anti-symmetric $\rep{10}$ of the structure group of $\Vt$.  Therefore the
number of massless chiral superfields in each representation, before
taking the quotient by $\IZ_{12}$, is given by
\begin{equation*}
   \begin{array}{l l}
      n_{\rep{10}} = h^1(\Xt, \Vt)~, & n_{\conjrep{5}} = h^1(\Xt, \wedge^2 \Vt) \\[1ex]
      n_{\conjrep{10}} = h^1(\Xt,\dual{\Vt})~, & n_{\rep{5}} = h^1(\Xt, \wedge^2 \dual{\Vt}) ~,
   \end{array}
\end{equation*}
where, for example, $h^i(\Xt,\Vt) = \dim_\IC H^i(\Xt,\Vt)$.  So it is these
cohomology groups we wish to calculate.  This will involve a
considerable amount of algebraic gymnastics, so although we try to
give complete arguments, many intermediate calculations are
relegated to \aref{app:cohomology}.

It will also be important to keep track of the $\IZ_{12}$ action on
the various cohomology groups.  We ultimately wish to break
$SU(5)_\GUT$ to $\GSM$ by turning on discrete Wilson lines, and
the massless fields on $X$ are those which are invariant under the
combined action of $\IZ_{12}$ on the cohomology groups, and the
Wilson lines on the corresponding $SU(5)_\GUT$ representations.
Note that an equivariant bundle map induces an equivariant map
between cohomology groups, so all our long exact sequences
actually split up into twelve independent sequences---one for each
irreducible $\IZ_{12}$ representation.

\subsection{Wilson line breaking of \texorpdfstring{$SU(5)_\GUT$}{SU(5) GUT}}\label{sec:Wilson_lines}

To break $SU(5)_\GUT$ down to $\GSM$, we turn on discrete
$SU(5)_\GUT$-valued holonomy around non-contractible paths in
$X$, such that the centraliser of this discrete holonomy group is
precisely $\GSM$.  This amounts to choosing a homomorphism from
$\pi_1(X) \cong \IZ_{12}$ to $SU(5)_\GUT$, so we must classify the
suitable choices.

The largest subgroup of $SU(5)_\GUT$ which commutes with $\GSM$
is the hypercharge group $U(1)_Y$, the embedding of which
is\footnote{We use the mathematically natural normalisation of
hypercharge; conventional hypercharge assignments are a factor of
three (or sometimes six) smaller.}
\begin{equation}\label{eq:hypercharge}
    U(1)_Y \ni~ e^{\ii\th} ~\mapsto~ \diag(e^{-2\ii\th},e^{-2\ii\th},e^{-2\ii\th},e^{3\ii\th},e^{3\ii\th}) ~.
\end{equation}
For appropriate symmetry breaking to occur, the image of $\IZ_{12}$
must lie in this subgroup, so there are eleven possible choices for the
Wilson lines, given by $g_{12} \mapsto e^{\frac{\pi\ii}{6}k} \in U(1)_Y$,
where $k = 1,\ldots, 11$.

Standard model matter resides in the $\rep{10}$ and $\conjrep{5}$ of
$SU(5)_\GUT$, which are the rank-two anti-symmetric tensor and the
anti-fundamental representation, respectively.  Since the symmetry is
broken by the discrete Wilson lines, different components of these
representations will carry different $\IZ_{12}$ `charges', which are
collected in \tref{tab:charges}.  Massless fields on $X$ come from
cohomology classes whose $\IZ_{12}$ transformation is conjugate
to that under the Wilson line, so that they combine to form an invariant.
\begin{table}[ht]
\begin{center}
\begin{tabular}{| l | c | c | c |c | c |}
    \hline\str
        Field  & $u^c$  & $Q$  & $e^c$  & $d^c$  & $L, H_d$ \\
    \hline\str
        $SU(5)$ provenance~  & $\rep{10}$  & $\rep{10}$  & $\rep{10}$  & $\conjrep{5}$  & $\conjrep{5}$ \\
    \hline\str
        $\GSM$ rep.  & $(\conjrep{3}, \rep{1})_{-4}$  & $(\rep{3},\rep{2})_{1}$  & $(\rep{1}, \rep{1})_{6}$
            & $(\conjrep{3},\rep{1})_2$  & $(\rep{1}, \rep{2})_{-3}$ \\
    \hline\str
        $\IZ_{12}$ charge  & $8k$  & $k$  & $6k$  & $2k$  & $9k$ \\
    \hline
\end{tabular}
\parbox{.8\textwidth}{\caption{\label{tab:charges}
\small
The standard model matter representations, their origin in $SU(5)_\GUT$
representations, and their `charges' under the discrete $\IZ_{12}$-valued Wilson
line.}}
\end{center}
\end{table}

We pause here briefly to clear up a possible point of confusion.  In
\sref{sec:Wilson_1} we explained that choosing a non-trivial equivariant
structure on the undeformed bundle
$\Vt_0 = T\Xt\oplus\cO_{\Xt}\oplus\cO_{\Xt}$ was equivalent to turning on
discrete Wilson lines on $X$ which commute with the $SU(3)$ holonomy
group of the tangent bundle, and therefore must lie in $E_6$.  In this
section, we have claimed that we are free to choose discrete Wilson
line values in $SU(5)_\GUT \subset E_6$.  Of course, we can only
choose the value of the Wilson lines once, so why is this allowed?

Note that the Wilson lines of \sref{sec:Wilson_1} commute with
$SU(5)_\GUT$, and therefore with the hypercharge group
\eqref{eq:hypercharge}.  So the full picture is that, before deforming
the bundle, the discrete Wilson lines lie in $E_6$ and are given by the
product of the element specified by $n_1, n_2$, as described in
\sref{sec:Wilson_1} and the element of $U(1)_Y$ specified by $k$,
described in this section.  This has the effect of breaking $E_6$ to
some smaller rank-six group such as\footnote{Given $n_1, n_2$ and $k$,
it is not too difficult to calculate the unbroken gauge group explicitly, but
this is somewhat tedious, and adds nothing to the current work.}
$SU(4)\times SU(2)\times U(1)^2$ or $SU(3)\times SU(2)\times U(1)^3$;
the possibilities were described long ago in \cite{Witten:1985xc}.  When
we deform to an irreducible rank-five bundle, this is physically equivalent
to Higgsing this extended gauge group to $\GSM$.  Mathematically, the
piece of the Wilson line which lies in $SU(5)$ gets mixed up with the
continuous holonomy group of the bundle, whereas the piece in
$U(1)_Y$ breaks $SU(5)_\GUT$ to $\GSM$.  In this way, contact is made
between our techniques and the older literature about heterotic standard
embedding models, although we emphasise again that our top down
approach of directly constructing the bundles is much more powerful
than effective field theory arguments.

\subsection{Exotic states from \texorpdfstring{$\rep{10}\oplus\conjrep{10}$}{10 + 10-bar}}
\label{sec:10s}

We will begin by calculating $H^1(\Xt, \Vt)$, corresponding to massless
chiral multiplets in the $\rep{10}$ of $SU(5)_\GUT$.  Since $h^i(\Xt, \cO_{\Xt}) = 0$
for $i = 1,2$, the long-exact cohomology sequence following from the
leftmost column of \eqref{eq:V_diagram} yields in part
\begin{equation*}
    0 \longrightarrow H^1(\Xt, \Ft) \longrightarrow H^1(\Xt, \Vt) \longrightarrow 0~,
\end{equation*}
so $H^1(\Xt, \Vt) \cong H^1(\Xt, \Ft)$, and we need to calculate the
latter group.  This fits into a long exact sequence following from the
middle row of \eqref{eq:V_diagram}, the relevant part of which is
\begin{equation*}
    \begin{array}{r c l c l c l c}
    0 & \longrightarrow & H^0(\Xt, \Ft) & \longrightarrow & H^0\big(\Xt, \bigoplus_{i=1}^{12} \cO_{\Xt}(D_i)\big) &
        \longrightarrow & H^0\big(\Xt, \cO_{\Xt}(\sum_i D_i)\big) & \longrightarrow \\[2ex]
        & \longrightarrow & H^1(\Xt, \Ft) & \longrightarrow & H^1\big(\Xt, \bigoplus_{i=1}^{12} \cO_{\Xt}(D_i)\big) &
        \longrightarrow & \ldots & 
    \end{array} \raisebox{-5ex}{~.}
\end{equation*}
Using $H^0(\Xt, \Ft) \cong \IC^6$ and results from \aref{app:cohomology},
this becomes
\begin{equation*}
    \begin{array}{r c c c l c l c}
    0 & \longrightarrow & \IC^6 & \longrightarrow & \IC^{12} &
        \longrightarrow & \IC^{48} & \longrightarrow \\[2ex]
        & \longrightarrow & H^1(\Xt, \Ft) & \longrightarrow & 0 &
        \longrightarrow & \ldots & 
    \end{array} \raisebox{-5ex}{~,}
\end{equation*}
where $\IC^{12} \sim \Reg_{\IZ_{12}}$ and
$\IC^{48} \sim 4*\Reg_{\IZ_{12}}$.  Our final result is therefore
\begin{equation*}
    H^1(\Xt, \Ft) ~\cong~ \IC^{42} \sim 3*\Reg_{\IZ_{12}}\oplus H^0(\Xt, \Ft) ~,
\end{equation*}
and we remind the reader that we are free to choose $H^0(\Xt, \Ft)$ to
transform as any sub-representation of \eqref{eq:H08O_rep}.  Finally,
we recall that the equivariant structure on $\Vt$ had to be twisted by
the $\IZ_{12}$ representation $\rep{\tilde n}$, where $\tilde n$ is
determined by \eqref{eq:ntilde}, so this extra phase acts on the
cohomology as well.  Taking into account that
$\tilde n \otimes\Reg_{\IZ_{12}} = \Reg_{\IZ_{12}}$ for any $\tilde n$,
we get
\begin{equation}\label{eq:H1V}
    H^1(\Xt, \Vt) \sim 3*\Reg_{\IZ_{12}}\oplus \big(\rep{\tilde n} \otimes H^0(\Xt, \Ft)\big) ~.
\end{equation}

We wish to find models with massless fields filling out precisely three
copies of the $\rep{10}$.  The regular representation, $\Reg_{\IZ_{12}}$,
contains each irreducible representation of $\IZ_{12}$ exactly once, so
\emph{any} choice of Wilson lines will lead to three massless copies of
the $\rep{10}$ coming from $3*\Reg_{\IZ_{12}}$.  We therefore ask that
no states originating in $H^0(\Xt, \Ft)$ survive the projection.  Referring
to \tref{tab:charges}, we see that this means choosing $k$ such that
$(\rep{k}\oplus\rep{6k}\oplus\rep{8k})\otimes \rep{\tilde n}\otimes H^0(\Xt, \Ft)
\supset\hspace{-.9em}/ ~\rep{0}$.

Multiplets transforming under the
$\conjrep{10}$ representation of $SU(5)_\GUT$ come from the
cohomology group $H^1(\Xt, \dual{\Vt})$, but there is no need to
calculate this independently, as we will now explain.  First we note
that Serre duality on $\Xt$ implies an isomorphism
$H^1(\Xt, \dual{\Vt}) \cong H^2(\Xt, \Vt)^*$, where the superscript
${}^*$ on the right-hand side indicates the dual vector space, and the latter
is easy to calculate.  Indeed, since $H^0(\Xt, \Vt) = H^3(\Xt,\Vt) = 0$ (the
vanishing of $H^3$ is demonstrated explicitly in \aref{app:stability}), the
Euler characteristic of $\Vt$ is simply
\begin{equation*}
    \chi(\Vt) = h^2(\Xt, \Vt) - h^1(\Xt, \Vt) ~,
\end{equation*}
and we know that $\chi(\Vt) = -36$, since $\Vt$ is a deformation of
$T\Xt{\oplus}\cO_{\Xt}{\oplus}\cO_{\Xt}$, and $\chi$ does not change
under deformation.
So $h^2(\Xt,\Vt)$ is determined indirectly by knowing $h^1(\Xt,\Vt)$
and $\chi(\Vt)$.  Furthermore, since $\Vt$ is equivariant under the
fixed-point-free group action, $\chi$ actually admits a simple
refinement.  For any irreducible representation $\rep{n}$ of $\IZ_{12}$,
let $h^i(\Xt, \Vt)_{\rep{n}}$ be the number of times this representation
appears in the decomposition of $H^i(\Xt, \Vt)$.  Then\footnote{In the
more general case of a group $G$ and a representation $\rep{R}$, the
fraction $\frac{1}{12}$ here is replaced by $\frac{\dim \rep{R}}{|G|}$.}
\begin{equation*}
    h^2(\Xt, \Vt)_{\rep{n}} - h^1(\Xt, \Vt)_{\rep{n}} = \frac{1}{12}\, \chi(\Vt) = -3 ~.
\end{equation*}
So if a particular representation occurs $k$ times in $H^1(\Xt, \Vt)$,
it necessarily occurs $k-3$ times in $H^2(\Xt, \Vt)$.  This means that
the \emph{conjugate} representation occurs $k-3$ times in
$H^1(\Xt, \dual{\Vt})$, since Serre duality involves a vector space
dualisation.  These fields transform in the $\conjrep{10}$ of $SU(5)_\GUT$,
so the Wilson lines also act on them in the conjugate representation.
Altogether, then, the fields projected out of $H^1(\Xt, \dual{\Vt})$ when
taking the quotient are precisely the conjugates of those projected out of
$H^1(\Xt, \Vt)$, so any massless fields extraneous to the three generations
of $\rep{10}$ will occur in vector-like pairs.

The final result is that there are 43 models
which have no exotic states originating in the $\rep{10}\oplus\conjrep{10}$
of $SU(5)_\GUT$.  The values of $n_1, n_2$ and $k$ for these models are
given in \tref{tab:43_models}.
\begin{table}[hbt]
\begin{center}
   \begin{tabular}{| c | c |}
      \hline
      ~$(n_1\, ,\, n_2)$~ & $k$ \\
      \hline\hline
      $(0\, , 3)$ & 2, 4, 8, 10 \\
      \hline
      $(0\, , 6)$ & 6 \\
      \hline
      $(0\, , 9)$ & 2, 4, 8, 10 \\
      \hline
      $(2\, , 3)$ & 6, 10 \\
      \hline
      $(2\, , 8)$ & 3, 6, 9 \\
      \hline
      $(2\, , 9)$ & 6, 10 \\
      \hline
      $(3\, , 4)$ & 4, 6, 7, 8, 10 \\
      \hline
      $(3\, , 8)$ & ~2, 4, 6, 8, 11~ \\
      \hline
      $(3\, , 10)$ & 2, 6 \\
      \hline
      $(4\, , 9)$ & 1, 4, 6, 8, 10 \\
      \hline
      $(4\, , 10)$ & 3, 6, 9 \\
      \hline
      $(8\, , 9)$ & 2, 4, 5, 6, 8 \\
      \hline
      $(9\, , 10)$ & 2, 6 \\
      \hline
   \end{tabular}\\
   \parbox{.85\textwidth}{\caption{\label{tab:43_models}\small
   A list of the 43 models, specified by the three discrete bundle parameters
   $n_1, n_2$ and $k$, which have no exotic massless fields descending
   from the $\conjrep{10}$ of $SU(5)_\GUT$.}
   }
\end{center}
\end{table}
\subsection{Doublet-triplet splitting}

To find the massless states coming from the $\rep{5}$ and $\conjrep{5}$
of $SU(5)_\GUT$, we must calculate the cohomology group
$H^1(\Xt,\wedge^2\Vt)$, and the $\IZ_{12}$ representation which
acts on it.  It will prove more convenient to instead calculate
$H^1(\Xt, \wedge^2\dual{\Vt})$, and apply the arguments of the last section
to relate one to the other, but this is still significantly more involved than
our previous calculations.  One tactic we will use repeatedly is to dualise
bundles and appeal to Serre duality, in order to move as much non-trivial
cohomology as possible into $H^0$ i.e. global sections.  These are much
easier to work with, as they are represented simply by homogeneous
polynomials, perhaps evaluated modulo some ideal.

We start with the short exact sequences which form the middle column and
bottom row of \eqref{eq:V_diagram}, defining respectively the bundles
$\Gt$ on $Z$, and $\Vt$ on the hypersurface $\Xt$.  Dualising these
yields\footnote{Note that, although $\dual{\cO_Z\!} \cong \cO_Z$, the
dualisation has the effect of conjugating the equivariant structure.  This will
be important when we are calculating the group action on cohomology.}
\begin{align}
    & 0 \longrightarrow \dual{\Gt} \longrightarrow \bigoplus_{i=1}^{12} \cO_Z(- D_i)
        \longrightarrow 6\cO_Z  \longrightarrow 0 ~, \label{eq:Gd_seq} \\[1ex]
    & 0 \longrightarrow \cO_{\Xt}(-\sum_{i=1}^{12} D_i) \longrightarrow \dual{\Gt}\rest{\Xt}
        \longrightarrow \dual{\Vt} \longrightarrow 0 ~. \label{eq:Vd_seq}
\end{align}
To get a handle on $\wedge^2 \dual{\Vt}$, we need something more.  For
any short exact sequence
\begin{equation*}
    0 \longrightarrow A \stackrel{\psi_1}{\longrightarrow} B
        \stackrel{\psi_2}{\longrightarrow} C \longrightarrow 0 ~,
\end{equation*}
there are associated exact sequences
\begin{align}
    0 \longrightarrow \wedge^2 A \longrightarrow \wedge^2 B \longrightarrow
        B\otimes C \longrightarrow S^2 C \longrightarrow 0 ~, \label{eq:W2_seq} \\[1ex]
    0 \longrightarrow S^2 A \longrightarrow A\otimes B \longrightarrow \wedge^2 B
        \longrightarrow \wedge^2 C \longrightarrow 0 ~, \label{eq:S2_seq}
\end{align}
where $S^2$ denotes the second symmetric power of a bundle, and the maps
are constructed in the obvious ways from $\psi_1$ and $\psi_2$.  So from
\eqref{eq:Vd_seq}, we can find an exact sequence for $\wedge^2\dual{\Vt}$,
\begin{equation*}
    0 \longrightarrow \cO_{\Xt}(-2\sum_{i=1}^{12} D_i) \longrightarrow
        \dual{\Gt}(-\sum_{i=1}^{12} D_i)\rest{\Xt} \longrightarrow \wedge^2 \dual{\Gt}\rest{\Xt}
         \longrightarrow \wedge^2 \dual{\Vt} \longrightarrow 0 ~.
\end{equation*}
If we introduce a bundle $K_1$, defined as the kernel of the map
$\wedge^2\dual{\Gt}\rest{\Xt} \to \wedge^2\dual{\Vt}$ (equivalently,
the image of the previous map), this splits into two short exact sequences
\begin{align}
    & 0 \longrightarrow \cO_{\Xt}(-2\sum_{i=1}^{12} D_i) \longrightarrow
        \dual{\Gt}(-\sum_{i=1}^{12} D_i)\rest{\Xt} \longrightarrow K_1 \longrightarrow 0 ~, \label{eq:K1_seq} \\[1.5ex]
    & \hspace{3em} 0 \longrightarrow K_1 \longrightarrow \wedge^2 \dual{\Gt}\rest{\Xt}
         \longrightarrow \wedge^2 \dual{\Vt} \longrightarrow 0 ~. \label{eq:W2Vd_seq}
\end{align}
In line with our general approach of considering only non-zero $H^0$ if
possible, we will in fact consider the dual of \eqref{eq:K1_seq},
\begin{equation*}
    0 \longrightarrow \dual{K_1\!} \longrightarrow \Gt(\sum_{i=1}^{12} D_i)\rest{\Xt}
        \longrightarrow \cO_{\Xt}(2\sum_{i=1}^{12} D_i) \longrightarrow 0 ~.
\end{equation*}
We know from \aref{app:X_cohomology} that the second and third bundles
here have $H^i = 0$ for $i>0$, so we immediately get
$H^i(\Xt, \dual{K_1\!}) = 0$ for $i=2,3$.  The other two groups fit into an
exact sequence
\begin{equation*}
    0 \longrightarrow H^0(\Xt, \dual{K_1\!}) \longrightarrow \IC^{372} \longrightarrow
        \IC^{312} \longrightarrow H^1(\Xt, \dual{K_1\!}) \longrightarrow 0 ~.
\end{equation*}
The middle map here is a complicated one, induced by $\Phi$, between
large vector spaces of polynomials reduced modulo the defining
polynomial of our hypersurface.  Nevertheless, it is possible to show
using computer algebra that this map is surjective, so that in particular,
$H^1(\Xt, \dual{K_1\!}) = 0$.  So now we have
$H^1(\Xt, \dual{K_1\!}) = H^2(\Xt, \dual{K_1\!}) = 0$, and Serre duality
tells us that the same groups vanish for $K_1$.  This information combines
with the long exact cohomology sequence from \eqref{eq:W2Vd_seq} to
give us the simple result
\begin{equation*}
    H^1(\Xt, \wedge^2\dual{\Vt}) ~\cong~ H^1(\Xt, \wedge^2\dual{\Gt}) ~.
\end{equation*}
We now have to compute this latter group.  First we will relate it to
cohomology on the ambient space $Z$, which can be calculated directly.
The short exact sequence relating $\wedge^2\dual{\Gt}\rest{\Xt}$ to
$\wedge^2\dual{\Gt}$ is
\begin{equation}\label{eq:W2Gd_rest_seq}
    0 \longrightarrow \wedge^2\dual{\Gt}\big(-\sum_{i=1}^{12} D_i\big) \longrightarrow
        \wedge^2\dual{\Gt} \longrightarrow \wedge^2\dual{\Gt}\rest{\Xt} \longrightarrow 0 ~.
\end{equation}
In \aref{app:Z_cohomology}, we show that
$H^i\big(Z, \wedge^2\dual{\Gt}(-\sum_i D_i)\big) = 0$ for $i<4$.
Plugging this into the long exact sequence following from
\eqref{eq:W2Gd_rest_seq} gives us the simple result that
$H^1(\Xt, \wedge^2\dual{\Gt}) = H^1(Z, \wedge^2\dual{\Gt})$, so we
have succeeded in lifting the required group to the ambient space.

Now return to \eqref{eq:Gd_seq}, which defines $\dual{\Gt}$ in terms of
line bundles.  Applying the general result \eqref{eq:S2_seq} to this
case yields another four-term exact sequence, which we again split into
two short exact sequences by introducing a kernel $K_2$,
\begin{align}
    & 0 \longrightarrow \wedge^2\dual{\Gt} \longrightarrow \bigoplus_{i<j} \cO_Z(-D_i -D_j)
        \longrightarrow K_2 \longrightarrow 0 ~, \label{eq:W2Gd_seq} \\[1ex]
    &\hspace{1em} 0 \longrightarrow K_2 \longrightarrow 6\bigoplus_i \cO_Z(-D_i) \longrightarrow
        21\cO_Z \longrightarrow 0 ~. \label{eq:K2_seq}
\end{align}
The middle term in \eqref{eq:K2_seq} has vanishing cohomology,
and the third term has $H^i = 0$ for $i>0$, so we immediately get
$H^i(Z, K_2) = 0$ for $i\neq1$, and
\begin{equation*}
    H^1(Z, K_2) ~\cong~ H^0(Z, 21\cO_Z) ~\cong~ \IC^{21} ~.
\end{equation*}
We can also take the cohomology of the middle term of
\eqref{eq:W2Gd_seq} from \aref{app:cohomology}; the only important
result is $H^1 \cong \IC^{18}$.  So the long exact cohomology
sequence coming from \eqref{eq:W2Gd_seq} reads, in part,
\begin{equation}\label{eq:crux}
    0 \longrightarrow H^1(Z, \wedge^2\dual{\Gt}) \longrightarrow \IC^{18} \longrightarrow
        \IC^{21} \longrightarrow \ldots ~~~.
\end{equation}
So to summarise, we have shown that $H^1(\Xt, \wedge^2 \Vt)$ is
isomorphic to the kernel of the map from $\IC^{18}$ to $\IC^{21}$
above, where these two groups are in fact respectively
$H^1\big(Z, \bigoplus_{i<j} \cO_Z(-D_i -D_j)\big)$ and $H^1(Z, K_2)
\cong H^0(Z, 21\cO_Z)$.  Finding this kernel is a tedious calculation
in \v{C}ech cohomology, and we will omit the details, except to note
that the task is simplified by the fact that these cohomology groups
are all generated by cocycles which are invariant under the torus
action, and are therefore represented by constant sections on each open
patch.

Of course, we also need to keep track of the $\IZ_{12}$
representations, noting that since the equivariant structure on $\Vt$
is twisted by $\rep{\tilde n}$, the action on the cohomology of
$\wedge^2 \Vt$ gets tensored by $\rep{2\tilde n}$.

Referring to \aref{app:cohomology}, we see that the generators of
$H^1\big(Z, \bigoplus_{i<j} \cO_Z(-D_i -D_j)\big) \cong \IC^{18}$
come from eighteen distinct summands of the bundle.
Specifically, we have $H^1\big(Z, \cO_Z(-D_1 -D_3)\big) \cong \IC$,
and similarly for the eleven other line bundles related to this one by
the $\IZ_{12}$ action.  The rest of the cohomology group originates in
$H^1\big(Z, \cO_Z(-D_1-D_4)\big) \cong \IC$, and the five other
bundles related to this one by $\IZ_{12}$.

So the twelve bundles related to $\cO_Z(-D_1 -D_3)$ by $\IZ_{12}$
obviously contribute one copy of $\Reg_{\IZ_{12}}$ to the cohomology.
There is an extra subtlety associated with the bundles related to
$\cO_Z(-D_1 -D_4)$, because the order-two element of the group maps
$\cO_Z(-D_1 -D_4)$ to itself (since $g_{12}^6 : D_1 \leftrightarrow D_4$),
and might do so with or without a minus sign.  There are two things to
consider in deciding which sign occurs.  Firstly, note that the bundle we
are discussing, $\bigoplus_{i<j} \cO_Z(-D_i -D_j)$, arises as the second
anti-symmetric power of $\bigoplus_i \cO_Z(D_i)$, so there is a minus
sign associated with exchanging the two factors of
$\cO_Z(-D_1 -D_4) = \cO_Z(-D_1) \otimes \cO_Z(-D_4)$.  Secondly, we
write down an explicit cocycle representing the generator of
$H^1\big(Z, \cO_Z(-D_1 - D_4)\big)$, act on it with $g_{12}^6$, and ask
whether the result is cohomologous to what we started with or its
negative.  It turns out that a minus sign occurs here as well, so overall,
$g_{12}^6$ acts trivially on $H^1\big(Z, \cO_Z(-D_1 - D_4)\big)$.
Therefore the $\IZ_{12}$ action on the six summands of this type factors
through $\IZ_6$, and in this way corresponds to $\Reg_{\IZ_6} \cong
\rep{0}\oplus\rep{2}\oplus\rep{4}\oplus\rep{6}\oplus\rep{8}\oplus\rep{10}$.
So altogether,
\begin{equation}\label{eq:H1W2}
    H^1\big(Z, \bigoplus_{i<j}\cO_Z(-D_i - D_j)\big) \sim \Reg_{\IZ_{12}}\oplus\Reg_{\IZ_6} ~.
\end{equation}

To find the representation content of $H^1(Z, K_2) \cong H^0(Z, 21\cO_Z)$,
recall that $21\cO_Z$ occurs as the second symmetric power of $6\cO_Z$,
which appears in \eqref{eq:Gd_seq}.  The only subtlety is to remember
that this sequence arose as the \emph{dual} to the sequence in
\eqref{eq:V_diagram}.  Obviously $\cO_Z$ is its own dual, but we must
remember to dualise the action on the cohomology.  This gives
\begin{equation}\label{eq:H1K2}
    H^1(Z, K_2) \sim S^2\big(H^0(\Xt, \Ft)^*\big) ~.
\end{equation}
So we now have the representation content of the two relevant terms in
the sequence \eqref{eq:crux}, which allows us to split it into twelve
sequences, one for each $\IZ_{12}$ irrep.  This step must be done
separately for each choice of the bundle parameters $n_1, n_2$, but
if we are looking for the MSSM spectrum, we can restrict ourselves to
those which appear in \tref{tab:43_models}.

\subsection{Models with the MSSM spectrum}

Let us choose one model to discuss in detail (with hindsight, we choose
one which yields the correct spectrum).  Let $n_1 = 3$ and $n_2 = 4$.
Then from \eqref{eq:ntilde} we get $\tilde n = 1$, and from the discussion
in \sref{sec:deformations} and \sref{sec:equivariance}, $H^0(\Xt, \Ft) \sim
\rep{0}\oplus\rep{2}\oplus\rep{6}\oplus\rep{8}\oplus\rep{9}\oplus\rep{10}$.
Plugging this into \eqref{eq:H1V}, we find
\begin{equation*}
    H^1(\Xt, \Vt) \sim 3*\Reg_{\IZ_{12}}\oplus
        \rep{1}\oplus\rep{3}\oplus\rep{7}\oplus\rep{9}\oplus\rep{10}\oplus\rep{11}
\end{equation*}
We now ask which values of $k$ we can choose for the Wilson lines (see
\sref{sec:Wilson_lines}) such that we get exactly three copies of the
$\rep{10}$ from this cohomology group.  As an example, take $k = 4$.
Consulting \tref{tab:charges}, we see that the three different components
of the $\rep{10}$ then have $\IZ_{12}$ charges 8, 4 and 0.  Taking the
tensor product of $\rep{0}\oplus\rep{4}\oplus\rep{8}$ with
$\rep{1}\oplus\rep{3}\oplus\rep{7}\oplus\rep{9}\oplus\rep{10}\oplus\rep{11}$
does not yield any invariants, so for this choice, there are no extra
massless states originating in the $\rep{10}\oplus\conjrep{10}$ of
$SU(5)_\GUT$.

Now we turn to the doublet-triplet splitting.  From \eqref{eq:H1K2} and
$H^0(\Xt, \Ft)$ above, we find that
\begin{equation*}
    H^1(\Xt, K_2) \sim 3*\rep{0}\oplus\rep{1}\oplus 2*\rep{2}\oplus\rep{3}\oplus 3*\rep{4}
        \oplus\rep{5}\oplus 3*\rep{6}\oplus\rep{7}\oplus 3*\rep{8}\oplus\rep{9}
        \oplus 2*\rep{10} ~.
\end{equation*}
Group equivariance means that the map from $\IC^{18}$ to $\IC^{21}$
breaks up into blocks, one for each irreducible $\IZ_{12}$ representation.
For example, we know from \eqref{eq:H1W2} that regardless of our choice
of $n_1, n_2$, $\IC^{18}$ contains one copy of $\rep{11}$.  But in the
present case, $\IC^{21}$ contains no instance of this representation,
so the kernel of the map must contain exactly one $\rep{11}$.  For the
other representations, we must do explicit calculations, and we find
\begin{equation*}
    H^1(Z, \wedge^2 \dual{\Gt}) \sim \rep{10}\oplus\rep{11} ~.
\end{equation*}
Finally, noting that $2*(3+4) \equiv 2$ mod 12, we tensor this representation
with $\rep{2}$ to obtain
\begin{equation}\label{eq:5_ex}
    H^1(\Xt, \wedge^2\dual{\Vt}) \sim \rep{0}\oplus\rep{1} ~.
\end{equation}
Referring to \tref{tab:charges}, and recalling that the above corresponds
to states in the $\rep{5}$ rather than $\conjrep{5}$, we see that the
triplets will have $\IZ_{12}$ charge $-2k \equiv 10k$ under the Wilson
line, while the doublets will have charge $-9k \equiv 3k$.  For our
choice $k = 4$, then, this corresponds to the representations $\rep{4}$
and $\rep{0}$ respectively.  For the doublets, this $\rep{0}$ pairs up with
the $\rep{0}$ in \eqref{eq:5_ex} to give us an invariant, and therefore a single
massless up-type Higgs doublet (the down-type doublet is its partner
from $H^1(\Xt,\wedge^2\Vt)$, guaranteed by the index theorem), while
for the triplets, there is no such invariant.

We have shown that in the case $(n_1, n_2) = (3,4)$ and $k=4$, we
obtain a model with exactly the massless spectrum of the MSSM.  In
total, there are eight choices which work equally well; they are listed
in \tref{tab:models}.
\begin{table}[htb]
\begin{center}
\begin{tabular}{| c | c |}
    \hline
        $(n_1, n_2)$  & $k$ \\
    \hline\hline
        $(3,4)$ & $4, 8$ \\
    \hline
        $(3,8)$ & $4, 8$ \\
    \hline
        $(4,9)$ & $4, 8$ \\
    \hline
        $(8,9)$ & $4, 8$ \\
    \hline
\end{tabular}
\parbox{.85\textwidth}{
    \caption{\label{tab:models}\small
    The values of the discrete bundle parameters $n_1, n_2$, and
    corresponding Wilson line parameters $k$, which lead to models
    with exactly the light spectrum of the MSSM.}
    }
\end{center}
\end{table}

\section{Discussion and conclusion} \label{sec:conclusion}

This paper achieves the long-standing goal of obtaining the MSSM
spectrum from the heterotic string by deforming a three generation
standard embedding solution \cite{Witten:1985bz}.  The obvious
difference to more general heterotic compactifications is the absence
of M5-branes or a non-trivial gauge bundle in the hidden sector, as
the anomaly cancellation condition is saturated by the second Chern
class of the visible sector bundle.

It is interesting to ask whether one might be able to obtain the MSSM
spectrum by supersymmetric deformation of some other three generation
standard embedding solution(s).  In order to break $SU(5)_\GUT$ to
$\GSM$ with discrete Wilson lines, the Calabi-Yau manifold $Y$ must
have non-trivial fundamental group, and therefore be obtained as a
quotient $Y = \widetilde Y/G$ for some finite group $G$, while to yield
three generations, its topological Euler characteristic must satisfy
$\chi(Y) = \pm 6$.

What are the possible choices for $Y$?    In fact, there are relatively few
Calabi-Yau threefolds with non-trivial fundamental group, and very few
of these have $\chi = \pm 6$.  Here we have taken $\widetilde X/\IZ_{12}$,
where the Hodge numbers of $\widetilde X$ are $\hodgenos = (8,44)$.
This manifold admits a different free quotient by the order-twelve
non-Abelian group $\Dic_3$ with the same Hodge numbers, but as we
discuss in \aref{app:Dic3_models}, there are no MSSM models on this
quotient.  The mirror manifold also admits at least one free quotient with
$\hodgenos = (4,1)$ \cite{Braun:2009qy}, and this might admit MSSM
models, although these are probably equivalent to models on the
$(1,4)$ manifolds by $(0,2)$ mirror symmetry \cite{Kreuzer:2010ph}.

In \cite{Davies:2011fr}, a manifold was constructed with
$\hodgenos = (2,5)$ and fundamental group $\IZ_5$.  With this fundamental
group, $SU(5)_\GUT$ cannot be broken to $\GSM$ by Wilson lines.

The most famous three generation manifold is perhaps Yau's manifold
\cite{Yau1}, which is a $\IZ_3$ quotient of a simply-connected manifold with
Hodge numbers $\hodgenos = (14,23)$.  If we repeat the arguments of
\sref{sec:10s} for this example, it is easy to see that we cannot arrange for all
the $\conjrep{10}$ states to be projected out by the quotient.  This is because
of the relatively large value of $h^{1,1}$ for the covering space, which
means that the three irreducible representations of $\IZ_3$ each appear
several times, and Wilson lines cannot be chosen to project them all out.  The
same problem will occur for all other known three generation manifolds, as
they all have relatively small fundamental groups and/or large Hodge numbers
\cite{Candelas:2008wb,Davies:2011fr,Triadophilia}.

So it seems that, at least if we restrict ourselves to known Calabi-Yau
threefolds, the models we have found are the only way to obtain the
exact massless spectrum of the MSSM from deforming the heterotic
standard embedding.  Of course, this is by no means enough for a
\emph{realistic} model of particle physics.  In particular, we have not
yet attempted to calculate the Yukawa couplings or address the problems
of supersymmetry breaking and moduli stabilisation.

\section*{Acknowledgements}

R. Davies is supported by the Engineering and Physical Sciences Research
Council [grant number EP/H02672X/1].  R. Donagi acknowledges partial
support by NSF grants 0908487 and 0636606.

\newpage
\appendix

\section{Cohomology calculations} \label{app:cohomology}

The Calabi-Yau $\Xt$ is defined as a hypersurface in $Z$ by $f=0$,
where $f$ is a section of the anticanonical bundle $\cO_Z(-K_Z)
\cong \cO_Z(\sum_i D_i)$, where $\{D_i\}$ are the twelve toric divisors.
As such, for any holomorphic vector bundle $B$, defined on $Z$, we
have the following short exact sequence,
\begin{equation} \label{eq:rest_seq}
    0 \longrightarrow B\big(-\sum_i D_i\big) \stackrel{f}{\longrightarrow} B \longrightarrow
        B\rest{\Xt} \longrightarrow 0 ~.
\end{equation}
This allows us to `lift' the important cohomology calculations to $Z$,
where they are a lot simpler.

The fourfold $Z$ is in fact a product, $Z \cong S_1\times S_2$, where
each surface is isomorphic to the del Pezzo surface $\dP_6$.  There are
corresponding projections $\pi_l : Z \to S_l$, and all the relevant line
bundles on $Z$ can be written as $\pi_1^*\cL_1\otimes\pi_2^*\cL_2$.
In this case we have the K\"unneth formula
\begin{equation*}
    H^i(Z, \pi_1^*\cL_1\otimes\pi_2^*\cL_2) \cong \bigoplus_{j+k=i}
        H^j(S_1, \cL_1)\otimes H^k(S_2, \cL_2) ~.
\end{equation*}
Many of our calculations therefore boil down to line bundle cohomology
on $\dP_6$, which we discuss in \aref{app:dP6_cohomology}.
Cohomology of $\Gt$ and related bundles on $Z$ is then calculated
in \aref{app:Z_cohomology}.  Finally, these results are used in
\aref{app:X_cohomology} to calculate the required cohomology groups
defined on the hypersurface $\Xt$.  We will use the notation
$h^\bullet(M, B) = (h^0, \ldots, h^{\dim M})$ for the dimensions of the
cohomology groups of a bundle $B$ on a manifold $M$.

\subsection{Line bundle cohomology on \texorpdfstring{$\dP_6$}{dP6}}\label{app:dP6_cohomology}

The surface $S\cong\dP_6$ is toric, which makes it relatively easy to
calculate line bundle cohomology using the \v{C}ech approach.
Throughout this section, we will refer to the six vectors $\{\n_a\}$ from
\eqref{eq:dP6_vertices}, and the corresponding toric divisors $\cD_a$.
We will also use $N$ and $M$ to refer to the two-dimensional lattices
relevant for $S$, rather than the four-dimensional lattices from
\sref{sec:manifold}.

Instead of thinking about line bundles in terms of local patches and
transition functions, it is clearer to utilise the divisor-line bundle
correspondence.  So for a toric divisor $\cD = \sum_{a=1}^6 c_a \cD_a$,
a section of $\cO_S(\cD)$ over an open patch $\cU$ is
a meromorphic function $f$, defined on
$\cU$, such that $(f) + \cD\rest{\cU} \geq 0$.

Each point $u$ of the lattice $M$ corresponds to an irreducible
character of the torus $(\IC^*)^2$, as well as a meromorphic function
$f_u$ on $S$, which transforms under the torus action according to
the corresponding character.  For any cone $\s$ in the fan for
$\dP_6$, the sections of $\cD$ over the corresponding open set are
given by
\begin{equation}\label{eq:toric_sections}
    \G\big(\cU_{\s}, \cO_S(\sum_{a=1}^6 c_a \cD_a)\big) = 
        \big\langle f_u ~\vert~ u\in M ~,~ \langle u, \n_a \rangle \geq -c_a
        ~\forall~ \n_a \in \s \big\rangle ~.
\end{equation}
In this way, the cohomology of any toric line bundle is graded by $M$.
Note that global sections are particularly simple; since they come from
points $u \in M$ satisfying the above in all open sets, $f_u$ contributes
to $H^0\big(S, \cO_S(\sum_{a=1}^6 c_a \cD_a)\big)$ if and only if
\begin{equation}\label{eq:inequalities}
    \langle u, \n_a\rangle \geq -c_a ~~\forall~~ a ~.
\end{equation}

Finally, note that the fan for $S$ has the symmetry of the hexagon,
$D_6$, and the surface inherits a faithful action of this group.  This
means that any bundles related by a $D_6$ transformation will have
the same cohomology groups (the toric weights will be related by the
corresponding dual transformation of the lattice $M$).

The rest of this section presents the cohomology of line bundles on
$S$ which are needed in \sref{sec:spectrum}.
Throughout this section, the indices $a, b,\ldots$ will run from 1 to
6, and arithmetical operations on them will be understood modulo 6.
We will take an open cover of $S$ consisting of six sets $\cU_a$
corresponding to the six two-dimensional cones in \fref{fig:dP6_fan}.

\subsubsection{Explicit calculations}

We will present in detail the calculation of the cohomology of one
particular line bundle (or class of line bundles, related by symmetry),
and then simply list the other results we need.\\[2ex]
\underline{$\cO_S(-\cD_a - \cD_{a+2})$}\\[1ex]
To be explicit, we will consider $\cO_S(-\cD_1 - \cD_3)$.  It is easy
to convince oneself that this has no global sections, so $H^0 = 0$.
To find $H^2$, we Serre dualise:
\begin{equation*}
    H^2\big(S, \cO_S(-\cD_1 - \cD_3)\big) \cong
        H^0\big(S, \cO_S(\cD_1 + \cD_3 - \sum_a \cD_a)\big)^* ~.
\end{equation*}
Again, it is not hard to see that the group on the right is zero by
considering the inequalities \eqref{eq:inequalities}.  So the only
cohomology group which might be non-vanishing is $H^1$,
and in this case we have simply $h^1 = -\chi$, where $\chi$ is
the Euler characteristic of the bundle.  This can be calculated from
the Hirzebruch-Riemann-Roch formula, which gives
$\chi = -1$.  So $H^1\big(S, \cO_S(-\cD_1 - \cD_3)\big) \cong \IC$.

In many cases, knowing the dimension of cohomology groups is
enough, but the hardest calculation in this paper is finding the
kernel of the map in \eqref{eq:crux}, and for this we need explicit
representatives for the cohomology classes.  Some of these
live in the $H^1$ groups of the type we have just calculated, so
our task is not complete.

First, we ask which points of the lattice $M$ might contribute to
$H^1$.  Notice from \eqref{eq:toric_sections} that the contribution
of weight $u\in M$ to the \v{C}ech cochain groups depends only
on whether $\langle u, \n_a\rangle \geq 0$ or
$\langle u, \n_a\rangle < 0$, for each value of $a$.  This divides
the lattice $M$ into various `chambers', with the contribution of
$u \in M$ to the cohomology depending only on the chamber to
which it belongs.  Since $S$ is compact, all cohomology groups
must be finite, so we need only consider chambers containing a
finite number of points.  In this case there turns out to be only
one of these, and the unique point it contains is the origin,
$0 \in M$, which corresponds to locally constant sections.

Referring again to \eqref{eq:toric_sections}, we can write down
explicit parametrisations of the \v{C}ech cochain groups
$\check{C}^0$ and $\check{C}^1$ with weight $0 \in M$,
\begin{align*}
    \check{C}^0\big(\cO_S(-\cD_1 - \cD_3)\big) &= 
        \{ (0,0,0,\a _1,\a _2,0) ~\vert~ \a_i \in \IC \} ~, \\[2ex]
    \check{C}^1\big(\cO_S(-\cD_1 - \cD_3)\big) &= 
        \{ \b _1,\b _2,\b _3,\b _4,0,0,\b _5,\b _6,\b _7,\b _8,\b _9,\b _{10},\b _{11},\b _{12},\b _{13}
        ~\vert~ \b_i \in \IC \} ~,
\end{align*}
where the open sets $\cU_a$ are ordered in the obvious way,
and double overlaps $\cU_{a,b} = \cU_a\cap \cU_b$ are ordered
first by $a$ and then by $b$.  Acting on the latter group with the
\v{C}ech differential $d_1$, we find a three-dimensional kernel.
Obviously the image of $d_0$ is two-dimensional (since
$\check{C}^0$ has two parameters $\a_1, \a_2$, and $H^0 = 0$),
so as expected, we find a one-dimensional cohomology group,
the generator of which can be taken to be the class of the cocycle
\begin{equation*}
    (-1,-1,0,0,0,0,1,1,1,1,1,1,0,0,0) ~\in~ \check{C}^1\big(\cO_S(-\cD_1 - \cD_3)\big)~.
\end{equation*}
By symmetry, we have
$h^\bullet\big(S, \cO_S(- D_a - D_{a+2})\big) = (0,1,0)$ for any
value of $a$, and we can obtain representatives for the cohomology
generators by acting on the one above with the appropriate
permutation (recalling that there is a sign change associated with
changing $\cU_{a,b}$ to $\cU_{b,a}$).

Other required cohomology dimensions (some of these results
follow from others by Serre duality) are:
\vspace{-2ex}
\begin{align*}
    h^\bullet(S, \cO_S) = (1,0,0)~,&
         ~~h^\bullet\big(S, \cO_S(-D_a)\big) = (0,0,0) ~,& \\[2ex]
    h^\bullet\big(S, \cO_S(- D_a - D_{a+1})\big) = (0,0,0) ~,&
         ~~h^\bullet\big(S, \cO_S(- D_a - D_{a+3})\big) = (0,1,0)   ~,& \\[2ex]
    h^\bullet\big(S, \cO_S(D_a - \sum_b D_b)\big) = (0,0,0)~,&
         ~~ h^\bullet\big(S, \cO_S(-\sum_a D_a)\big) = (0,0,1) ~,& \\
    h^\bullet\big(S, \cO_S(D_a)\big) = (1,0,0) ~,&
        ~~h^\bullet\big(S, \cO_S(D_a + D_{a+1})\big) = (2,0,0) ~,& \\[2ex]
    h^\bullet\big(S, \cO_S(D_a + D_{a+2})\big) = (1,0,0) ~,&
        ~~h^\bullet\big(S, \cO_S(D_a + D_{a+3})\big) = (1,0,0) ~,& \\[2ex]
    h^\bullet\big(S, \cO_S(\sum_a D_a - D_b)\big) = (5,0,0) ~, &
        ~~ h^\bullet\big(S, \cO_S(\sum_a D_a)\big) = (7,0,0) ~,& \\[1ex]
    ~~ h^\bullet\big(S, \cO_S(D_a + \sum_b D_b)\big) = (8,0,0) ~,&
        ~~ h^\bullet\big(S, \cO_S(2\sum_a D_a)\big) = (19,0,0) ~.
\end{align*}

\newpage
\subsection{Cohomology on \texorpdfstring{$Z$}{Z}}\label{app:Z_cohomology}

The cohomology of line bundles on $Z$ follows immediately from
our calculations on $S \cong \dP_6$ above, by the K\"unneth
formula.  But we also need to know the cohomology of the
bundle $\Gt$ and related bundles.  We will make use of Serre
duality to reduce the number of calculations we need to do,
remembering that the canonical class of $Z$ is given by
$K_Z \sim -\sum_i D_i$.\\[3ex]
\underline{$\Gt$ and $\dual{\Gt}(-\sum_i D_i)$}\\[1ex]
The bundle $\Gt$ is defined by the short exact sequence
\begin{equation*}
    0 \longrightarrow 6\cO_Z \longrightarrow \bigoplus_i \cO_Z(D_i)
        \longrightarrow \Gt \longrightarrow 0 ~.
\end{equation*}
Using the results of \aref{app:dP6_cohomology} and the K\"unneth
formula, the resulting long exact sequence in cohomology breaks
up into several pieces
\begin{align*}
    &0 \longrightarrow \IC^6 \longrightarrow \IC^{12} \longrightarrow H^0(Z, \Gt) \longrightarrow 0 ~,\\[2ex]
    &\hspace{2em}0 \longrightarrow H^i(Z, \Gt) \longrightarrow 0 ~,\quad i>0 ~,
\end{align*}
giving the simple result
\begin{equation*}
    h^\bullet(Z, \Gt) = (6,0,0,0,0) ~,
\end{equation*}
and hence, by Serre duality,
\begin{equation*}
    h^\bullet\big(Z, \dual{\Gt}(-\sum_i D_i)\big) = (0,0,0,0,6) ~.
\end{equation*}

\noindent\underline{$\Gt(\sum_i D_i)$}\\[1ex]
Twisting the sequence which defines $\Gt$ by $\cO_Z(\sum_i D_i)$
yields another exact sequence
\begin{equation*}
    0 \longrightarrow 6\cO_Z(\sum_i D_i) \longrightarrow \bigoplus_i \cO_Z(D_i + \sum_j D_j)
        \longrightarrow \Gt(\sum_i D_i) \longrightarrow 0 ~.
\end{equation*}
Once again, the associated long exact sequence breaks up very
simply,
\begin{align*}
    &0 \longrightarrow \IC^{294} \longrightarrow \IC^{672} \longrightarrow H^0\big(Z, \Gt(\sum_i D_i)\big)
        \longrightarrow 0 ~,\\[2ex]
    &\hspace{3em}0 \longrightarrow H^i\big(Z, \Gt(\sum_i D_i)\big) \longrightarrow 0 ~,
\end{align*}
and we find
\begin{equation*}
    h^\bullet\big(Z, \Gt(\sum_i D_i)\big) = (378,0,0,0,0) ~.
\end{equation*}

\newpage
\noindent\underline{$\wedge^2\Gt$ and $\wedge^2\dual{\Gt}(-\sum_i D_i)$}\\[1ex]
Consider the exact sequences obtained by dualising
\eqref{eq:W2Gd_seq} and \eqref{eq:K2_seq}
\begin{align*}
    &0 \longrightarrow \dual{K_2\!} \longrightarrow \bigoplus_{i<j}\cO_Z(D_i + D_j)
        \longrightarrow \wedge^2\Gt \longrightarrow 0~,\\[2ex]
    &\hspace{.7em}0 \longrightarrow 21\cO_Z \longrightarrow 6\bigoplus_i \cO_Z(D_i) \longrightarrow
        \dual{K_2\!} \longrightarrow 0~.
\end{align*}
As in the previous cases, the long exact sequence from the second
of these is very simple, yielding
$h^\bullet(Z, \dual{K_2\!}) = (51,0,0,0,0)$.  The middle term of the
first sequence here is a direct sum of sixty-six line bundles.
Referring to \sref{app:dP6_cohomology}, we see that twelve of
them have $h^0 = 2$, the rest have $h^0 = 1$, while all have
vanishing higher cohomology.  So altogether, the first sequence
gives
\begin{align*}
    &0 \longrightarrow \IC^{51} \longrightarrow \IC^{78} \longrightarrow
        H^0(Z, \wedge^2\Gt) \longrightarrow 0 ~,\\[1ex]
    &\hspace{2em}0 \longrightarrow H^i(Z, \wedge^2\Gt) \longrightarrow 0 ~,\quad i>0 ~.
\end{align*}
Our final result is therefore
\begin{equation*}
    h^\bullet(Z, \wedge^2\Gt) = (27,0,0,0,0) ~~,~~ h^\bullet\big(Z, \wedge^2\dual{\Gt}(-\sum_i D_i)\big) = (0,0,0,0,27) ~.
\end{equation*}
\subsection{Cohomology on \texorpdfstring{$\Xt$}{X-tilde}}\label{app:X_cohomology}

The key to calculating the cohomology of bundles on $\Xt$ is the
exact sequence \eqref{eq:rest_seq}, since most of the bundles of
interest arise as restrictions of bundles on $Z$.  There are several
calculations needed for the arguments of Sections
\ref{sec:deformations} and \ref{sec:spectrum}.\\[3ex]
\underline{$\cO_{\Xt}(D_i)$}\\[1ex]
From \eqref{eq:rest_seq}, we get the exact sequence
\begin{equation*}
    0 \longrightarrow \cO_Z(D_i - \sum_j D_j) \longrightarrow \cO_Z(D_i)
       \longrightarrow  \cO_{\Xt}(D_i) \longrightarrow 0 ~.
\end{equation*}
The results of \aref{app:dP6_cohomology} show that the first term here
has vanishing cohomology, so the cohomology groups of
$\cO_{\Xt}(D_i)$ are the same as those of the corresponding bundle
on $Z$,
\begin{equation*}
    h^\bullet\big(\Xt, \cO_{\Xt}(D_i)\big) = (1,0,0,0) ~.
\end{equation*}

\noindent\underline{$\cO_{\Xt}(\sum_i D_i - D_j)$}\\[1ex]
This time, \eqref{eq:rest_seq} becomes
\begin{equation*}
    0 \longrightarrow \cO_Z(-D_j) \longrightarrow \cO_Z(\sum_i D_i - D_j)
        \longrightarrow \cO_{\Xt}(\sum_i D_i - D_j) \longrightarrow 0 ~.
\end{equation*}
From the results of \aref{app:dP6_cohomology}, we see that the first
bundle has zero cohomology, and the second only has non-vanishing
$H^0$; in fact,
\begin{equation*}
    H^0\big(Z, \cO_Z(\sum_i D_i - D_j)\big) ~\cong~ \IC^{35} ~,
\end{equation*}
so we immediately get
\begin{equation*}
    h^\bullet\big(\Xt, \cO_{\Xt}(\sum_i D_i - D_j)\big) = (35, 0,0,0) ~.
\end{equation*}

\noindent\underline{$\cO_{\Xt}(\sum_i D_i)$}\\[1ex]
In this case, \eqref{eq:rest_seq} gives
\begin{equation*}
    0 \longrightarrow \cO_Z \longrightarrow \cO_Z(\sum_i D_i)
        \longrightarrow \cO_{\Xt}(\sum_i D_i) \longrightarrow 0 ~.
\end{equation*}
The results of \aref{app:dP6_cohomology} show that all higher
cohomology groups vanish for the first two bundles here, and
\begin{equation*}
    H^0(Z, \cO_Z) ~\cong~ \IC ~~,~~
        H^0\big(Z, \cO_Z(\sum_i D_i)\big) ~\cong~ \IC^{49}~,
\end{equation*}
so we get
\begin{equation*}
    h^\bullet\big(\Xt, \cO_{\Xt}(\sum_i D_i)\big) = (48, 0,0,0) ~.
\end{equation*}

\noindent\underline{$\cO_{\Xt}(2\sum_i D_i)$}\\[1ex]
The relevant exact sequence here is
\begin{equation*}
    0 \longrightarrow \cO_Z(\sum_i D_i) \longrightarrow \cO_Z(2\sum_i D_i)
        \longrightarrow \cO_{\Xt}(2\sum_i D_i) \longrightarrow 0 ~.
\end{equation*}
This is very similar to the last example; we have
\begin{equation*}
    H^0\big(Z, \cO_Z(\sum_i D_i)\big) ~\cong~ \IC^{49} ~~,~~
        H^0\big(Z, \cO_Z(2\sum_i D_i)\big) ~\cong~ \IC^{361}~,
\end{equation*}
and all higher cohomology of both bundles vanishes.  We therefore
immediately obtain
\begin{equation*}
    h^\bullet\big(\Xt, \cO_{\Xt}(2\sum_i D_i)\big) = (312, 0,0,0) ~.
\end{equation*}

\noindent\underline{$\Ft$ and $\dual{\Ft}$}\\[1ex]
In \sref{sec:deformations} we discussed that by choosing the map
$\Phi$ appropriately, we could arrange that $H^0(\Xt, \Ft) \cong \IC^6$
and $H^1(\Xt, \Ft) \cong \IC^{42}$, but did not discuss the higher
cohomology groups.  The results above allow us to find these missing
groups, and by Serre duality, those of $\dual{\Ft}$.
We have the exact sequence
\begin{equation*}
    0 \longrightarrow \Ft \longrightarrow \bigoplus_i \cO_{\Xt}(D_i) \longrightarrow
        \cO_{\Xt}(\sum_i D_i) \longrightarrow 0 ~.
\end{equation*}
We saw above that the second and third bundles have vanishing
higher cohomology, which immediately implies that
$H^i(\Xt, \Ft) = 0$ for $i=2,3$, and hence by Serre duality that
$H^i(\Xt, \dual{\Ft}) = 0$ for $i=0,1$.  So in particular,
$H^0(\Xt, \dual{\Ft}) = 0$, which we used in \sref{sec:deformations} to show
that $\Ft_0$ is a completely non-split extension.
\vskip3ex

\noindent\underline{$\Gt(\sum_i D_i)\rest{\Xt}$}\\[1ex]
In this case we have
\begin{equation*}
    0 \longrightarrow \Gt \longrightarrow \Gt(\sum_i D_i) \longrightarrow
        \Gt(\sum_i D_i)\rest{\Xt} \longrightarrow 0 ~.
\end{equation*}
Referring to \aref{app:Z_cohomology}, we see that the first two
bundles again only have non-vanishing $H^0$, so the same
is true for the third; in fact we find
\begin{equation*}
    h^\bullet\big(\Xt, \Gt(\sum_i D_i)\rest{\Xt}\big) = (372,0,0,0) ~.
\end{equation*}
\section{A simple proof of stability} \label{app:stability}

In \sref{sec:bundle}, we outlined one argument for the stability of a generic bundle
$\Vt$ in our family on $\Xt$, following the work of Li and Yau.  Here we give a simple,
direct argument for stability of the corresponding (still generic) bundle $V$ on the
quotient space $X$.  By the Donaldson-Uhlenbeck-Yau (DUY) theorem on $X$, this
implies the existence of a Hermitian-Yang-Mills (HYM) connection on $V$. The
pullback to $\Xt$ is a HYM connection on $\Vt$, and by the DUY theorem again, this
time on $\Xt$, it follows that $\Vt$ is in fact also polystable.

The reason it is easier to work on $X$ is that its Picard group is cyclic, i.e.
$h^{1,1}(X)=1$ (and $h^{1,0}(X) = 0$). This allows us to invoke a special case of
Hoppe's theorem \cite{Anderson:2007nc,Anderson:2008ex}:  Let $X$ be a Calabi-Yau
threefold with $h^{1,1}(X) = 1$, and $V$ a holomorphic vector bundle on $X$ with
$c_1(V) = 0$.  If $H^0(X, \wedge^pV) = 0$ for $p = 1,\ldots,rk(V)-1$, then $V$ is stable.
One fact which makes our life easier is that since $\wedge^5 V \cong \cO_X$, we have
$\wedge^p V \cong \dual{\wedge^{5-p}V}$, and therefore by Serre duality,
$H^0(X, \wedge^p V) \cong H^3(X, \wedge^{5-p} V)^*$.

To demonstrate that our rank-five bundles $V$ are stable, it is therefore sufficient to
show that $H^0(X, V) = H^3(X, V) = 0$, and $H^0(X, \wedge^p V) = 0$ for $p=2,3$.
To do so, we will show that the corresponding cohomology groups of $\Vt$ already
vanish on the covering space $\Xt$; to this end, consider the representation of $\Vt$
via the short exact sequence \eqref{eq:seq_F}, which we repeat here
\begin{equation*}
    0 \longrightarrow 6\cO_{\Xt} \longrightarrow \Ft \longrightarrow \Vt \longrightarrow 0 ~.\tag{\ref{eq:seq_F}}
\end{equation*}
We have already explained in \sref{sec:deformations} that we can arrange for
$H^0(\Xt, \Vt) = 0$.  In \sref{app:X_cohomology}, we saw that $H^3(\Xt, \cF) =  0$,
and if we use this in the long exact cohomology sequence following from above,
we immediately get $H^3(\Xt, \Vt) = 0$, which by Serre duality implies
$H^0(\Xt,\wedge^4 \Vt) = 0$.

Calculating $H^0(\Xt, \wedge^2 \Vt)$ explicitly from the exact sequences is much
harder, but we can make a simple argument that it vanishes.  Using the techniques
we have described, we could instead construct irreducible rank-\emph{four} bundles
$\Vt^{(4)}$ as deformations of $\Vt_0^{(4)} = T\Xt{\oplus}\cO_{\Xt}$.  We then have
$\wedge^2 \Vt_0^{(4)} = \wedge^2 T\Xt{\oplus}T\Xt$, and hence
$h^0(\Xt, \wedge^2 \Vt_0^{(4)}) = 0$.  The semi-continuity theorem for sheaf
cohomology implies that at a general point in moduli space,
$h^0(\Xt, \wedge^2 \Vt^{(4)})$ is bounded above by this value \cite{Hartshorne}, and
it must therefore also vanish.  Now, we can consider our rank five bundles $\Vt$ to
be deformations of $\Vt^{(4)}{\oplus}\cO_{\Xt}$ (simply go to a point in moduli space
where the map $\Phi$ annihilates seven of the Euler vectors, instead of only six),
and we repeat the argument:  as long as $\Vt^{(4)}$ and $\wedge^2 \Vt^{(4)}$ have
no global sections, the same is true for $\wedge^2\big(\Vt^{(4)}{\oplus}\cO_{\Xt}\big)$,
and by the semi-continuity theorem, also for $\wedge^2\Vt$ at a general point in
moduli space.

A similar argument gives $H^0(\Xt, \wedge^3 \Vt) = 0$ as long as $H^0(\Xt, \Vt) = 0$,
so this is sufficient to satisfy the conditions for Hoppe's theorem, and therefore
conclude that $V$ is stable on the quotient space.

\section{The non-Abelian quotient} \label{app:Dic3_models}

The covering manifold $\Xt$ also admits a free quotient by the non-Abelian
dicyclic group $\Dic_3$, which yields another manifold with Hodge numbers
$\hodgenos = (1,4)$.  Here we will explain briefly why this manifold does not
yield MSSM models in the same way that the $\IZ_{12}$ quotient does.  For
a much more detailed discussion of $\Xt/\Dic_3$, see \cite{Braun:2009qy}.

First we need to understand the group $\Dic_3$ and its representations.  It can
be generated by two elements, one of order three and one of order four,
satisfying the additional relation
\begin{equation*}
    g_4\,g_3\, g_4^{-1} = g_3^2 ~,
\end{equation*}
which reveals a semi-direct produce structure
$\Dic_3 \cong \IZ_3 \rtimes \IZ_4$.  There are therefore four distinct
one-dimensional representations, in which $g_3$ acts trivially and $g_4$
corresponds to multiplication by one of the fourth roots of unity.  As in
\cite{Braun:2009qy}, we will denote
these representations by $R_1, R_{-1}, R_{\ii}, R_{-\ii}$.  There are also two
two-dimensional irreps, which can be distinguished by $\Tr(g_4^2) = \pm 2$;
these we denote by $R^{(2)}_\pm$.

The choice of discrete Wilson line values to break $SU(5)_\GUT$ to $\GSM$
is much more restricted than in the $\IZ_{12}$ case, since there are now only
three non-trivial one-dimensional representations.  The corresponding Wilson
lines are
\begin{equation}\label{eq:Dic3_Wilson}
    g_4 \mapsto \diag(1,1,1,-1,-1) ~,~ g_4 \mapsto \diag(-1,-1,-1,\ii,\ii)
        ~,~ g_4 \mapsto \diag(-1,-1,-1,-\ii,-\ii) ~,
\end{equation}
and in every case, $g_3$ is trivial.  We note a simple feature here.  Since the
$\rep{10}$ of $SU(5)$ is the rank-two anti-symmetric tensor, in the second or
third cases here it contains three distinct one-dimensional representations of
$\Dic_3$.  In the first case, though, all the $\rep{10}$ fields transform as
either $R_1$ or $R_{-1}$.  We will return to this momentarily.

We know from the discussion in \sref{sec:spectrum} that the construction of an
equivariant stable deformation of $T\Xt{\oplus}\cO_{\Xt}{\oplus}\cO_{\Xt}$
involves choosing a two-dimensional subspace of $H^{1,1}(\Xt)$ which is
invariant under the quotient group.  After deforming, the corresponding
massless fields in the $\rep{10}\oplus\conjrep{10}$ of $SU(5)$ become
massive via the Higgs mechanism.  Under $\Dic_3$, we have
\begin{equation*}
    H^{1,1}(\Xt) \sim R_1\oplus R_{-1}\oplus R_{\ii}\oplus R_{-\ii}\oplus R^{(2)}_+
        \oplus R^{(2)}_+ ~.
\end{equation*}
We can therefore choose either $R^{(2)}_+$, or the sum of any two
distinct one-dimensional representations.  Either way, when we calculate
$H^1(\Xt, \dual{\Vt})$, we will find at least two distinct one-dimensional
representations of $\Dic_3$.  Now recall that the second and third choices
in \eqref{eq:Dic3_Wilson} lead to fields in the $\conjrep{10}$ transforming
under at least three distinct one-dimensional representations.  Therefore
in these cases we are guaranteed to have massless fields from
the $\conjrep{10}$ which survive the quotient.  However, if we choose
$R_{1}\oplus R_{-1}$, we are left with
\begin{equation*}
    H^1(\Xt, \dual \Vt) \sim R_{\ii}\oplus R_{-\ii}\oplus R^{(2)}_+ \oplus R^{(2)}_+ ~.
\end{equation*}
Combining these representations with the first choice of Wilson line in
\eqref{eq:Dic3_Wilson} gives no invariants, and therefore projects out all
massless states from the $\conjrep{10}$.

By the above reasoning, the analogue for the $\Dic_3$ quotient of the 43
models in \tref{tab:43_models} is a unique model, in which the gauge
bundle $V$ on $X' = \Xt/\Dic_3$ is a deformation of
\begin{equation*}
    \cL_{R_{-1}}\otimes\big( TX'{\oplus}\cL_{R_{-1}}{\oplus}\cL_{R_1}\big)~,
\end{equation*}
and the discrete Wilson line values are given by
\begin{equation*}
    g_3 \mapsto \id_5 ~,~ g_4 \mapsto \diag(1,1,1,-1,-1) ~\in~SU(5)_{\GUT} ~.
\end{equation*}
We can then proceed to calculate $H^1(\Xt, \wedge^2 \dual{\Vt})$ for this model,
and we find
\begin{equation*}
    H^1(\Xt, \wedge^2 \dual{\Vt}) \sim R_{-1}\oplus R_1 \oplus 2 R^{(2)}_- ~.
\end{equation*}
Combining this with the Wilson line, we see that although we find a massless
pair of Higgs doublets, they come along with massless colour triplets.
We conclude that there are no models on the $\Dic_3$ quotient which yield
exactly the massless spectrum of the MSSM.

\newpage

\bibliographystyle{utphys}
\bibliography{references}

\end{document}

%% file: macros.tex
\usepackage{amssymb,amsmath}

\renewcommand{\a}{\alpha}
\renewcommand{\b}{\beta}
\newcommand{\G}{\Gamma}
\newcommand{\D}{\Delta}

\renewcommand{\th}{\theta}

\renewcommand{\l}{\lambda}\renewcommand{\L}{\Lambda}
\newcommand{\m}{\mu}
\newcommand{\n}{\nu}

\newcommand{\s}{\sigma}
\renewcommand{\t}{\tau}

\renewcommand{\o}{\omega}\renewcommand{\O}{\Omega}

\newcommand{\cD}{\mathcal{D}}

\newcommand{\cF}{\mathcal{F}}
\newcommand{\cG}{\mathcal{G}}

\newcommand{\cL}{\mathcal{L}}
\newcommand{\cM}{\mathcal{M}}
\newcommand{\cN}{\mathcal{N}}
\newcommand{\cO}{\mathcal{O}}

\newcommand{\cU}{\mathcal{U}}

\newcommand{\IC}{\mathbb{C}}

\newcommand{\IP}{\mathbb{P}}

\newcommand{\IR}{\mathbb{R}}

\newcommand{\IZ}{\mathbb{Z}}

\newcommand{\rep}[1]{\mathbf{#1}}
\newcommand{\conjrep}[1]{\overline{\mathbf{#1}}}
\newcommand{\rest}[1]{\big\vert_{#1}}

\newcommand{\quotient}[1]{_{\hskip-2pt\lower1pt\hbox{$/$}\lower2pt\hbox{\hskip-1pt$#1$}}}
\newcommand{\ii}{\text{i}}

\def\place#1#2#3{\vbox to0pt{\kern-\parskip\kern-7pt
                             \kern-#2truein\hbox{\kern#1truein #3}
                             \vss}\nointerlineskip}

\newcommand{\sref}[1]{Section~\ref{#1}}
\newcommand{\aref}[1]{Appendix~\ref{#1}}
\newcommand{\fref}[1]{Figure~\ref{#1}}
\newcommand{\tref}[1]{Table~\ref{#1}}

\newcommand{\Ext}{\mathrm{Ext}}

\newcommand{\Dic}{\mathrm{Dic}}
\newcommand{\diag}{\mathrm{diag}}
\newcommand{\GUT}{\mathrm{GUT}}
\newcommand{\GSM}{G_{\mathrm{SM}}}
\newcommand{\Hom}{\mathrm{Hom}}
\newcommand{\Reg}{\mathrm{Reg}}

\newcommand{\Tr}{\mathrm{Tr}}

\newcommand{\dual}[1]{#1^*}

\newcommand{\hodgenos}{(h^{1,1},\,h^{2,1})}
\newcommand{\symm}[1]{_{\hskip-3pt\lower3pt\hbox{$\left\{#1\right\}$}}}

\newcommand{\cicystop}{~\lower8pt\hbox{.}}

\def\place#1#2#3{\vbox to0pt{\kern-\parskip\kern-7pt
                             \kern-#2truein\hbox{\kern#1truein #3}
                             \vss}\nointerlineskip}

\newcommand{\beq}{\begin{equation}}
\newcommand{\eeq}{\end{equation}}
\newcommand{\bea}{\begin{eqnarray}}
\newcommand{\eea}{\end{eqnarray}}
\newcommand{\bean}{\begin{eqnarray*}}
\newcommand{\eean}{\end{eqnarray*}}

\newcommand{\comment}[1]{}

\newcommand{\+}{\phantom{-}}
\newcommand{\id}{{\bf 1}}

\newcommand{\dP}{\text{dP}}